\documentclass[onecolumn]{IEEEtran}

\usepackage[english]{babel}
\usepackage{float}

\usepackage[latin1]{inputenc}
\usepackage{amsmath}
\usepackage{amsfonts}
\usepackage{amssymb}
\usepackage{epsfig}
\usepackage{graphicx}
\usepackage{epstopdf}
\usepackage{array}
\usepackage{url}
\usepackage[table]{xcolor}

\usepackage{colortbl}
\usepackage{color}
\usepackage{cite}

\usepackage{flushend}
\begin{document}

\title{Energy Evaluation of Preamble Sampling MAC Protocols for Wireless Sensor
Networks} 

\author{
 	\IEEEauthorblockN{Giorgio
Corbellini\IEEEauthorrefmark{1}\IEEEauthorrefmark{2}, Cedric
Abgrall\IEEEauthorrefmark{1}, Emilio Calvanese
Strinati\IEEEauthorrefmark{1}, and Andrzej Duda\IEEEauthorrefmark{2}}
 	\IEEEauthorblockA{\IEEEauthorrefmark{1}CEA-LETI, MINATEC, Grenoble, France}
 	\IEEEauthorblockA{\IEEEauthorrefmark{2}Grenoble Institute of Technology, CNRS
Grenoble Informatics Laboratory UMR 5217, France}

 	Email: [Giorgio.Corbellini, Cedric.Abgrall, Emilio.Calvanese-Strinati]@cea.fr,
Andrzej.Duda@imag.fr
}

\maketitle

\begin{abstract}

The paper presents a simple probabilistic analysis of the
  energy consumption in preamble sampling MAC protocols. We validate the
  analytical results with simulations.
We compare the classical MAC protocols (B-MAC and X-MAC) with LA-MAC, a method
proposed in a companion paper. 
Our analysis highlights the energy savings achievable with LA-MAC with respect
to B-MAC and X-MAC. It also shows that
LA-MAC provides the best performance in the considered case of high density
networks under traffic congestion.
\end{abstract}  

\section{Introduction}

Wireless Sensor Networks (WSN) have recently expanded to support diverse
applications in various and ubiquitous scenarios, especially in the context of
Machine-to-Machine (M2M) networks~\cite{exalted_project}.
Energy consumption is still the main design goal along with providing 
sufficient performance support for target applications.    
Medium Access Control (MAC) methods play the key role in reducing energy
consumption~\cite{langendoen08medium} because of the part taken by the radio in
the overall energy budget. 
Thus, the main goal consists in designing an access method that reduces the
effects of both \emph{idle listening} during which a device consumes energy while
waiting for an eventual transmission and \textit{overhearing} when it receives a
frame sent to another device~\cite{langendoen08medium}.    

To save energy, devices aim at achieving low duty cycles: they alternate long
sleeping periods (radio switched off) and short active ones (radio switched
on).
As a result, the challenge of MAC design is to synchronize the instants of the
receiver wake-up with possible transmissions of some devices so that the
network achieves a very low duty cycle.
The existing MAC methods basically use two approaches.
The first one synchronizes devices on a common sleep/wake-up schedule by
exchanging synchronization messages (SMAC~\cite{ye02energy},
TMAC~\cite{vandam03adaptive}) or defines a synchronized network wide TDMA
structure (LMAC~\cite{vanhoesel04lightweight}, D-MAC~\cite{dmac.lu.2004},
TRAMA~\cite{rajendran05energy}).
With the second approach, each device transmits before each data frame a
\emph{preamble} long enough to ensure that intended receivers wake up to catch
its frame (Aloha with Preamble Sampling~\cite{elhoiydi02aloha}, Cycled
Receiver~\cite{lin04power}, LPL (Low Power Listening) in
B-MAC~\cite{polastre04versatile}, B-MAC+~\cite{avvenuti.bmacplus.2006},
CSMA-MPS~\cite{mahlknecht04csma} aka X-MAC~\cite{buettner06xmac},
BOX-MAC~\cite{kuntz2011auto}, and DA-MAC~\cite{corb11damac}).
Both approaches converge to the same scheme, called \emph{synchronous preamble
sampling}, that uses very short preambles and requires tight synchronization
between devices (WiseMAC~\cite{enz04wisenet}, Scheduled Channel Polling
(SCP)~\cite{ye06ultra}).

Thanks to its lack of explicit synchronization, the second approach based on
\textit{preamble sampling} appears to be more easily applicable, more scalable,
and less energy demanding than the first synchronous approach.
Even if methods based on \textit{preamble sampling} are collision prone, they
have attracted great research interest, so that during last years many protocols
have been published.
In a companion paper, we have proposed LA-MAC, a Low-Latency
Asynchronous MAC protocol~\cite{corb11lamac} based on preamble sampling and
designed for efficient adaptation of device behaviour to varying network
conditions.

In this paper, we analytically and numerically compare
B-MAC~\cite{polastre04versatile}, X-MAC~\cite{buettner06xmac}, and LA-MAC in
terms of energy consumption.
The novelty of our analysis lies in how we relate
the expected energy consumption to traffic load.
In prior energy analyses, authors based the expected energy consumption on the
average Traffic Generation Rate (TGR) of devices~\cite{ye06ultra} as well as on
the probability of receiving a packet in a given interval~\cite{buettner06xmac}.
In contrast to these approaches, which only focus on the consumption of a
``transmitter-receiver'' couple, we rather consider the global energy cost of
a group of neighbour contending devices.
Our analysis includes the cost of all radio operations involved in
the transmission of data messages, namely the cost of transmitting, receiving,
idle listening, and overhearing.

The motivation for our approach comes from the fact that in complex, dense, and
multi-hop networks, traffic distribution is not uniformly spread over the
network.
Thus, the expected energy consumption depends on traffic pattern,  \textit{e.g.}
\textit{convergecast}, \textit{broadcast}, or \textit{multicast}, because
instantaneous traffic load may differ over the network.
In our approach, we estimate the expected energy consumption that depends on the
instantaneous traffic load in a given localized area.
As a result, our analysis estimates the energy consumption independently of
the traffic pattern.


\section{Background}
\label{sec_preliminaries}

We propose to evaluate the expected energy consumption of a group of
sensor nodes under
three different preamble sampling MAC protocols:
B-MAC, X-MAC and LA-MAC. 
In complex, dense, and multi-hop networks, the instantaneous traffic distribution
over the network is not uniform.
For example, in the case of networks with the \textit{convergecast} traffic
pattern (all messages must be delivered to one sink), the the traffic load is
higher at nodes that are closer to the sink in terms of number of hops. 
Due to this \textit{funnelling effect}~\cite{wan2005siphon}, devices close to
the sink exhaust their energy much faster than the others.

The evaluation of the expected energy consumption in this case is difficult and
the energy analyses published in the literature often base the expected energy
consumption of a given protocol on the traffic generation rate of the
network~\cite{ye06ultra}.
In our opinion, this approach does not fully reflect the complexity of the
problem, so we propose to analyze the expected consumption with
respect to the number of messages that are buffered in a given geographical
area.
This approach can simulate different congestion situations by varying the
instantaneous size of the buffer.

In our analysis, we consider a ``star'' network composed of a single receiving
device (\textit{sink}) and a group of $N$  devices that may have
data to send.
All devices are within 1-hop radio coverage of each other.
We assume that all transmitting devices share a global message buffer for which
$B$ sets the number of queued messages, $B$ is then related to network
congestion.
Among all $N$ devices, $N_s$ of them have at least one packet
to send and are called \textit{active} devices.
Remaining  devices have empty buffers and do not participate in the
contention, nevertheless, they are prone to the \textit{overhearing effect}.
Thus, there are $N_o=N-N_s$ \textit{overhearers}.
According to the global buffer state $B$, there are several combinations of how
to distribute $B$ packets among $N$ sending devices: depending on the
number of packets inside the local buffers of active devices, $N_s$ and $N_o$ may
vary for each combination.
For instance, there can be $B$ active devices with each one packet to send or
less than $B$ active devices with some of them having more than one buffered
packet.

In the remainder, we explicitly separate the energy cost due to transmission
$E_t$, reception $E_r$, polling (listening for any radio activity in the
channel) $E_l$, and sleeping $E_s$.
$E_o$ is the overall energy consumption of all overhearers.
The overall expected energy consumption $E$ is the sum of all these energies.
The power consumption of respective radio states is $P_t$, $P_r$, $P_l$, and
$P_s$ for transmission, reception, channel polling, and sleeping.
The power depends on a specific radio device.
We distinguish the polling state from the reception state.
When a node is performing channel polling, it listens to any channel for
activity---to be detected, a radio transmission must start after the beginning
of channel polling.
Once a radio activity is detected, the device immediately switches its radio
state from polling to receiving.
Otherwise, the device that is polling the channel cannot change its radio state.
The duration of a message over the air is $t_d$.
The time between two wakeup instants is $t_f=t_l+t_s$, where $t_l$ and $t_s$ are respectively the
channel polling duration and the sleep period.  These values are related to the
duty cycle.

\section{Preamble Sampling MAC Protocols}
\label{sec_MAC_protocols}

\begin{figure}[ht!]
	\begin{center} 	
	\includegraphics[width=0.7\linewidth,clip=true]{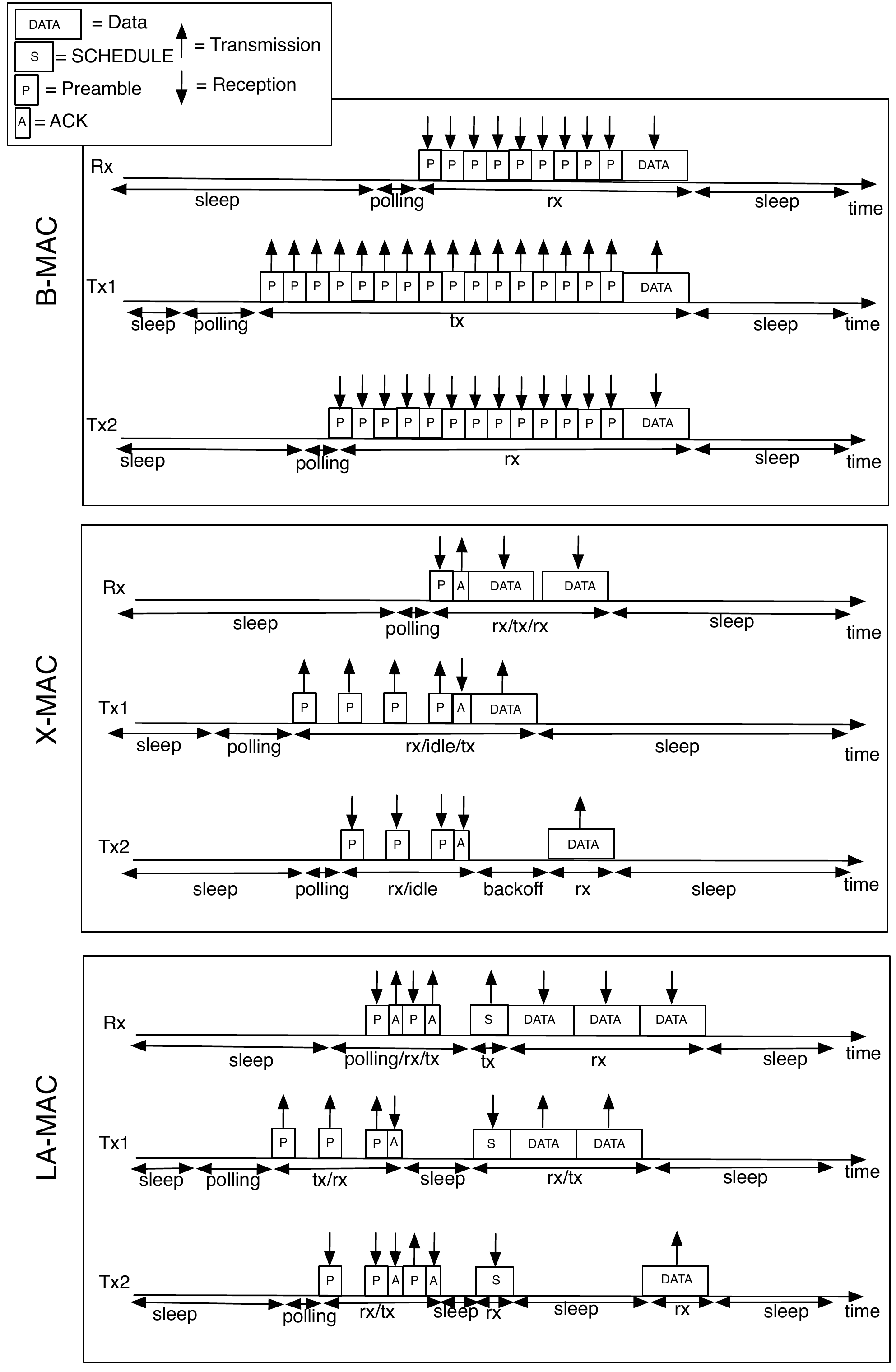}		
	\end{center} 	
	\caption{Comparison of analyzed MAC methods.} 
	\label{fig:operation_compare} 
\end{figure}

In this section, we provide the details of analyzed preamble sampling
protocols. 
Figure~\ref{fig:operation_compare} presnts the operation of all protocols.

\subsection{B-MAC}

In B-MAC \cite{polastre04versatile}, all nodes
periodically repeat the same cycle during their lifetime: wake up, listen for
the channel, and then go back to sleep.  
When an active node wants to transmit a
data frame, it first transmits a preamble long enough to cover the entire sleep
period of a potential receiver. 
After the preamble the sender immediately
transmits the data frame.  
When the receiver wakes up and detects the preamble,
it switches its radio to the receiving mode and listens to until the complete
reception of the data frame. 
Even if the lack of synchronization results in low
overhead, the method presents several drawbacks due to the length of the
preamble: high energy consumption of transmitters, high latency, and limited
throughput.
In the remainder, we define $t_p^B$, the duration of the B-MAC preamble.

\subsection{X-MAC}

In CSMA-MPS \cite{mahlknecht04csma} and X-MAC~\cite{buettner06xmac},  nodes
periodically alternate sleep and polling periods.
After the end of a polling period, each active node transmits a series
of short preambles spaced with gaps.
During a gap, the transmitter switches to the idle mode and expects to receive
an ACK from the receiver.
When a receiver wakes up and receives a preamble, it sends an ACK back to the
transmitter to stop the series of preambles, which reduces the energy spent by
the transmitter. 
After the reception of the ACK, the transmitter sends a data
frame and goes back to sleep.
After data reception, the receiver remains awake for possible transmission of a
single additional data frame.
If another active node receives a preamble destined to the same
receiver it
wishes to send to, it stops transmitting to listen to the channel for an
incoming
ACK.
When it overhears the ACK, it sets a random back-off timer at which
it
will send its data frame.
The transmission of a data frame after the back-off is not preceded by any
preamble.
Note however that nodes that periodically wake up to sample the channel
need to keep listening for a duration that is larger than the gap between short
preambles to be able to decide whether there is an ongoing transmission or not. 
The duration of each short preamble is $t_p^X$ and 
the ACK duration is $t_a^X$. 

\subsection{LA-MAC}

%

LA-MAC~\cite{corb11lamac} is a scalable protocol that aims at achieving
low latency and limited energy consumption by
building on three main ideas: efficient
forwarding based on proper scheduling of children nodes that want to transmit,
transmissions of frame bursts, and traffic differentiation.  
The method periodically adapts local organization of channel access depending on
network
dynamics such as the number of active users and the instantaneous traffic load.
In LA-MAC, nodes periodically alternate long sleep periods and short polling
phases.
During polling phases each receiver can collect several requests for
transmissions
that are included inside short preambles.
After the end of its polling period, the node that has collected some preambles
processes the requests, compares the priority of requests with the locally
backlogged messages and broadcasts a SCHEDULE message.
The goal of the SCHEDULE message is to temporarily organize the transmission of
neighbor nodes to avoid collisions.
If the node that ends its polling has not detected any channel activity and has
some
backlogged data to send, it starts sending a sequence of
short unicast
preambles containing the information about the burst to send.
As in B-MAC and X-MAC, the strobed sequence is long enough to wakeup the
receiver.
When a receiver wakes up and receives a preamble, it clears it with an ACK frame
containing
the instant of a \textit{rendezvous} at which it will broadcast the
SCHEDULE frame. 
If a second active node overhears a preamble destined to the same destination
it wants to send to, it waits
for an incoming ACK.
After ACK reception, a sender goes to sleep and wakes up at the instant
of the rendezvous.
In Figure~\ref{fig:operation_compare}, we see that after the transmission of an
ACK to Tx1,  Rx device is again
ready for receiving preambles from other devices.
So, Tx2 transmits a preamble and receives an ACK with the same rendezvous.
Preamble clearing continues until the end of the channel polling interval of
the receiver. 


\section{Energy Analysis}
\label{sec:energy_analysis}

LA- MAC provides its best performance in contexts of high density and traffic congestion. 
In order SHOW THE GAIN of LA-MAC, we provide an energy analysis aimed at comparing EXPECTED energy consumption of all considered protocols. 

We focus on evaluating expected energy consumption of a group of nodes when the number of messages to transmit within the group is known. 
In our analysis, we consider one receiver and a group of devices that can have some messages to send as well as empty buffers. 
In the analysis thta we provide, we focus our attention to the fact that in a complex sensor network traffic congestion is not uniformly distributed over the network.
In fact elements such as  the MAC protocol, the density and the traffic model have different impact in different areas of the network.
For this reason instead of focusing on the simple Traffic Generation Rate (TGR)~\cite{ye06ultra} on the probability of receiving a packet in a given interval~\cite{buettner06xmac}, we base our analysis on the number of messages that a group of nodes must send to a reference receiver.

With this approach we can show different congestion situations as they happen in a multi-hops networks with convergecast traffic pattern, where traffic distribution is not uniform with respect to proximity to the sink (in terms of number of hops). 
In fact, the closer the sink, the higher the average traffic.
We provide an evaluation that shows energy consumption with respect to a group of nodes.
We assume that a group of nodes share a global message buffer, depending on the number of messages in the buffer there may be zero, one, two or multiple senders.
Those nodes that have any message to send are called \textit{others} or \textit{overhearers}, they don't participate in the contention but are prone to the \textit{overhearing problem} (one of the major causes of energy waste in wireless sensor networks).
 
In the analysis we separate energy cost due to transmission (couple, triple or more) $E_t$, reception $E_r$,  polling (listening some activities in the channel) $E_l$ and  sleep $E_s$.  
Consumption of other node that overhears the channel is represented by $E_o$.  
Overall expected energy consumption $E$ is  the sum of all energies.
Global buffer state of the group of nodes is $B$. 
Power consumption of  radio states are $P_t$ for transmission, $P_r$ for reception, $P_l$ for channel polling and $P_s$ for sleep.
We assume that when a device is polling the channel, it listens to the air interface for some activity; if a message is already being sent while a device starts polling the channel, the device will not change its radio state. 
Otherwise, if a device that is polling the channel hears the beginning of new message, it switches its radio in receiving mode increasing the energy consumption.
We consider that the group is composed by $N$ devices and one receiver. 
Depending on the state of buffers, the number of senders $N_s$ varies as well as the number of overhearing nodes $N_o = N - N_s$. 
We assume that all  devices are within radio range of each others. 
Duration of a message over the air is $t_d$.
Each frame elapses $t_f = t_l + t_s$

\subsection{Global buffer is empty ($B=0$)}
If all buffers are empty,  all protocols behave in the same way:  nodes periodically wakeup, poll the channel, then go back to sleep because of absence of channel activity. Consumption only depends on time spent in polling  and sleeping.
\begin{equation}
E^{ALL}(0) = (N+1) \cdot ( t_l \cdot P_l+ t_s \cdot P_s)
\end{equation}

\subsection{Global buffer contains one message ($B=1$)}
If there is one message to send, there are only two devices that are active: the one which has a message in the buffer  ($N_s$ = 1) and the destination. 
The number of overhears is $N_o=N-1$.


\textit{ }

\noindent\textbf{B-MAC ($B=1$)}

When message sender wakes up, it polls the channel and then starts sending one large preamble that anticipates data  transmission. 
Even if data is unicast, destination field is not included in preambles; therefore, all nodes need to hear both preamble and the header of the following data in order to know the identity of the intended receiver. 
Provided that devices are not synchronized, each device will hear in average half of the preamble.  
The cost of transmission is the cost of an entire preamble plus the cost of transmitting data.
\begin{equation}
E_t^{B}(1) =   ( t_p^B  + t_d  ) \cdot  P_t
\end{equation}
The cost of reception is the cost of receiving half of the duration of a preamble plus the cost of receiving data.
In packetized radios, a large preamble is obtained by a sequence of short preambles sent one right after the other.
For this reason, if a generic device B wakes up and polls the channel while a generic device A is sending a long peamble, radio state of device B will remain in polling state for a short time until the beginning of the next small packet of the large preamble; afterwards the radio will switch in receiving mode consuming more energy. 
When the receiver (that is not synchronized with the sender)  wakes up, it polls the channel for some activity.
Because of lack of synchronization, it may happen that at the time when the receiver wakes up, the sender is performing channel polling. 
Probability of this event is $p=t_l/t_f$, so if the receiver wakes up during this period, it will perform half of the polling and then it will listen for the entire preamble. 
Otherwise, if the receiver wakes up after the end of the polling of sender, it will listen half of the preamble (probabiliy $1-p$).
In the remainder of this document we say that with probability $p$ transmitter and receiver are somehow quasi-synchronized. 
\begin{equation}
E_r^{B}(1) =  (p \cdot t_p^B + (1-p) \cdot \frac{t_p^B}{2} + t_d  ) \cdot  P_r
\end{equation}
So more than the entire polling of the sender we must consider half of polling period that must be performed by the receiver with probability $p$.
\begin{equation}
E_l^{B}(1) = (1 + \frac{p}{2})  \cdot t_l \cdot P_l  
\end{equation}
The cost of sleeping activity for the couple transmitter/receiver it depends on  the time that they do not spend in polling, receiving or transmitting messages.
\begin{equation}
E_s^{B}(1) = (2 \cdot t_f - ( \frac{t_p^B}{2} \cdot (p+3)  + 2 \cdot t_d  + t_l \cdot (1 + \frac{p}{2}) ))\cdot P_s 
\end{equation}
With B-MAC there is not difference in terms of energy consumption  between overhearing and receiving a message. Therefore, the cost of overhearing is: 
\begin{equation}
E_o^{B}(1)=  N_o\cdot (E_r^{B}(1) + p \cdot \frac{t_l}{2}   \cdot P_l  +  (t_f -  (p \cdot (\frac{t_l}{2} + t_p^B) + (1-p) \cdot \frac{t_p^B}{2} + t_d )) \cdot P_s ) 
\end{equation}


\textit{ }

\noindent\textbf{X-MAC ($B=1$)}

When the sender wakes up, it polls the channel and starts sending a sequence of unicast preambles separated by a time for \textit{early}  ACK reception.
When the intended receiver wakes up and polls the channel, it receives the preamble and clear it. 
Then the sender can transmit its message.
After data reception, the receiver remains in polling state for an extra backoff time $t_b$ that is used to receive other possible messages~\cite{buettner06xmac} coming from other senders. 
All devices that have no message to send, overhear channel activity and go to sleep as soon as they receive any unicast message (preamble, ACK or data). 
The expected number of preambles that are needed to \textit{wakeup} the receiver is $\gamma^X$. 
Average number of preambles depends on the duration of polling period, preamble and ACK messages as well as the duration of an entire frame~\cite{buettner06xmac}.
$\gamma^X$ Is the inverse of the \textit{collision} probability of one preamble over the polling period of the receiver.
In fact, if the couple sender/receiver is not synchronized, the sender can not know when the receiver will wake up, thus each preamble has the same probability to be heard or not by the receiver.
Each sent preamble  is a trial of a geometric distribution, so we say that before there is a collision between preamble and polling period there are $(\gamma^X-1)$ preambles whose energy is wasted.

\begin{equation}
\gamma^X =  \frac{1}{\frac{t_l - t_a^X - t_p^X}{t_f}} 
\end{equation}

Total amount of energy that is due to the activity of transmitting one message depends on the average number of preambles that must be sent ($\gamma^X $) and the cost of \textit{early} ACK reception. 
Provided that wakeup schedules of nodes are not synchronous, it may happen that  when the receiver wakes up, the sender is performing channel polling (transmitter and receiver are  quasi-synchronized with probability $p$). 

In the  case of quasi-synchronization, the receiver will perform in average half of the polling period and afterwards it the will be able to clear the very first preamble of the strobe.
In this case the cost of transmission only includes the transmission of one preamble and the cost of receiving the ACK.
Otherwise, if nodes are not synchronous (the receiver wakes up after the end of the polling of sender), the receiver will cause the sender to waste energy for the transmission of $\gamma^X$ preambles and the wait for an ACK (we consider waiting for ACK as a polling state) before it can hear one of them. 
The energy consumption of all activities of polling is reported separately in $E_l^{X}(1)$. 
Transmission cost is:
\begin{equation}
E_t^{X}(1) =  (1-p)\cdot \gamma^X  \cdot t_p^X \cdot P_t  + p \cdot t_p^X \cdot P_t  + t_a^X  \cdot P_r  +  t_d \cdot P_t 
\end{equation}

\begin{equation}
=  ((1-p)\cdot \gamma^X + p ) \cdot t_p^X \cdot P_t  + t_a^X  \cdot P_r  +  t_d \cdot P_t 
\end{equation}
The cost of the receiving activity is represented by the transmission of  one ACK and the reception of both data and preamble.

\begin{equation}
E_r^{X}(1) =   (t_d  + t_p^X ) \cdot  P_r +   t_a^X  \cdot P_t
\end{equation}

With probability $1-p$ (no synchronization)  the receiver will wakeup while the sender is already transmitting a preamble (or it is waiting for an \textit{early} ACK). 
Otherwise (with probability $p$) the receiver will perform in average, only half of its polling period.
The reason for this is that if the active couple is quasi-synchronized they simultaneously perform channel sensing, then the sender starts the preamble transmission. 
As far as the sender is concerned, we must consider both the entire polling period and the time that the sender waits for \textit{early} ACK without any answer (event that happens with probability  $1-p$).

\begin{equation}
E_l^{X}(1) =   ( ( t_l  +  (1-p)\cdot (\gamma^X - 1) \cdot t_a^X) + ((1-p)\cdot \frac{t_p^X + t_a^X}{2} +  p \cdot \frac{t_l}{2}) + t_b)\cdot  P_l
\end{equation}
\begin{equation}
 =   ((1-p)\cdot (\frac{t_p^X + t_a^X}{2} +(\gamma^X - 1) \cdot t_a^X )+  (\frac{p}{2}+  1)\cdot t_l + t_b)\cdot  P_l
\end{equation}
Sleep activity of the active couple is twice a frame duration minus the time that both devices are active. 

\begin{equation}
E_s^{X}(1) =   (2\cdot t_f - (t_l + ((1-p)\cdot \gamma^X + p) \cdot ( t_p^X + t_a^X) +  t_d ) - ( p \cdot \frac{t_l}{2}+t_p^X + t_a^X + (1-p)\cdot\frac{t_p^X + t_a^X}{2}  + t_d + t_b))\cdot P_s 
\end{equation}
\begin{equation}
=   (2\cdot t_f - 2 \cdot  t_d-  p \cdot \frac{t_l}{2}-t_p^X - t_a^X - (1-p)\cdot \frac{t_p^X + t_a^X}{2}  - t_l - ((1-p)\cdot \gamma^X + p) \cdot ( t_p^X + t_a^X)  - t_b)\cdot P_s 
\end{equation}

\begin{figure}[h!]
\begin{center}
\includegraphics[scale=.5]{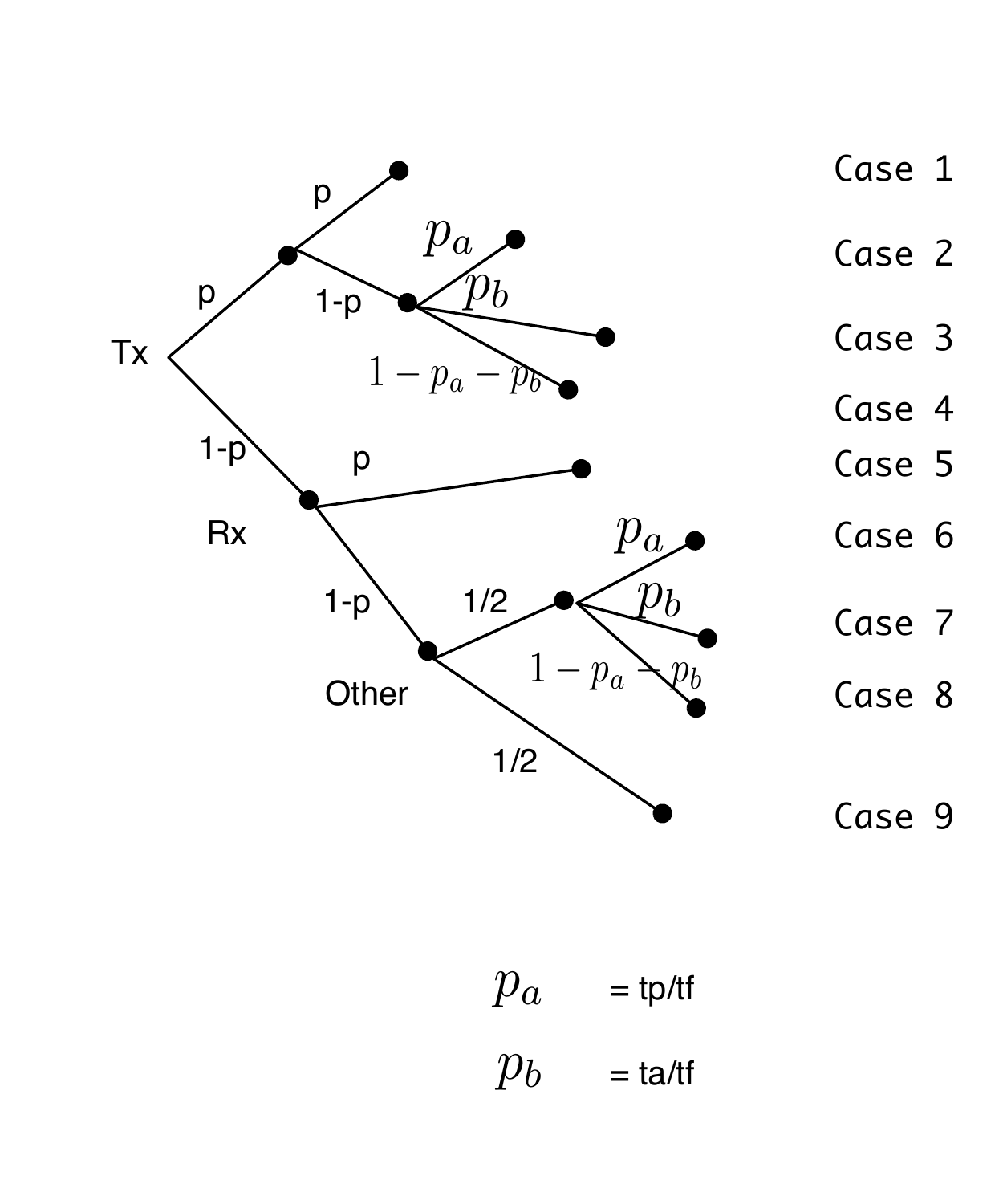}
\end{center}
\caption{X-MAC. Tree of different wakeup cases.} 
\label{fig:probabilitiesOthersXmac}
\end{figure}

As other devices, the overhearers can wakeup at a random instant. 
However, differently from active agents, as soon as they overhear some activity they go back to sleep. 
Therefore their energy consumption depends on the probability that such nodes wake up while the channel is busy or not. 
The probability that at wakeup instant the channel is free depends on polling duration,  buffer states, the number of senders etc.
In figure~\ref{fig:probabilitiesOthersXmac} we present all the possible situations that can happen. 
We consider as reference instant, the time at which the transmitter wakes up (root of the tree).
With probability $p$, the receiver and the the transmitter are quasi-synchronized, not synchronized otherwise (probability $(1-p)$).
With probability $p\cdot p$  both the receiver and a generic overhearer are quasi-synchronized with the transmitter, this is the Case 1 in the tree.

\vspace{0.2cm}
\vspace{0.2cm}

\begin{itemize}
\item Case 1 :  Sender, receiver and overhearer are quasi-synchronized. The  overhearer will sense a preamble that is not intended to it and the goes back to sleep. 
\vspace{1cm}
\begin{figure}[ht!]
\begin{center}
\includegraphics[scale=.5]{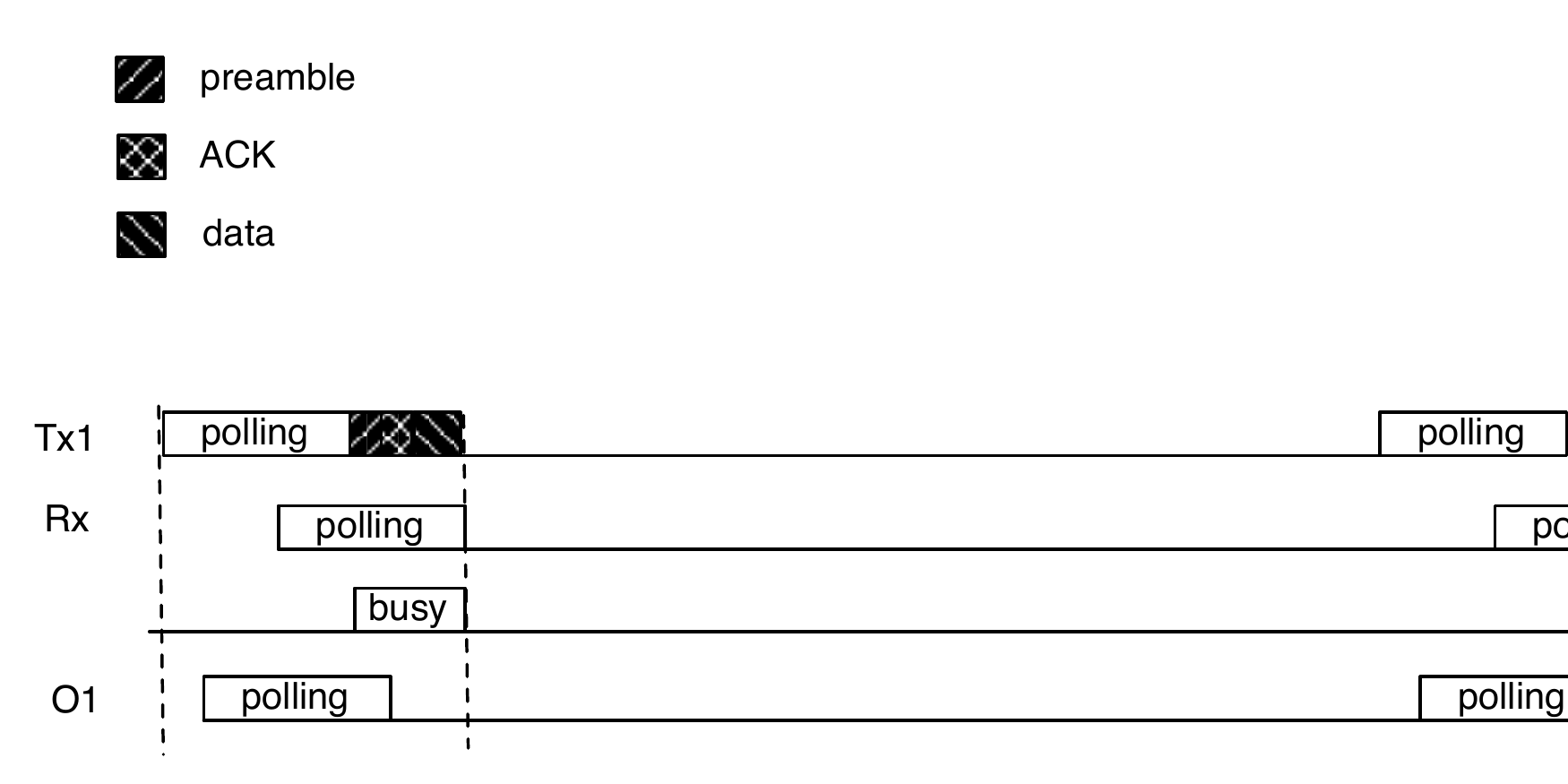}
\end{center}
\caption{Global buffer size A=1. Overhearing situations for Case 1. X-MAC protocol} 
\label{fig:xmacA1case1}
\end{figure}
\vspace{1cm}
\begin{equation}
	E_{Case_1,o}^X = \frac{t_l}{2} \cdot P_l + t_p^X \cdot P_r +  (t_f - \frac{t_l}{2} -t_p^X) \cdot P_s
\end{equation}
\item Case 2, 3, 4: Sender and receiver are synchronized but not the overhearer. 
When the overhearar  wakes up, it can ovehear different messages (preamble, ACK or data) as well as clear channel.
Possible situations are summarized in figure~\ref{fig:xmacA1case234}.
\vspace{1cm}
\begin{figure}[ht!]
\begin{center}
\includegraphics[scale=.5]{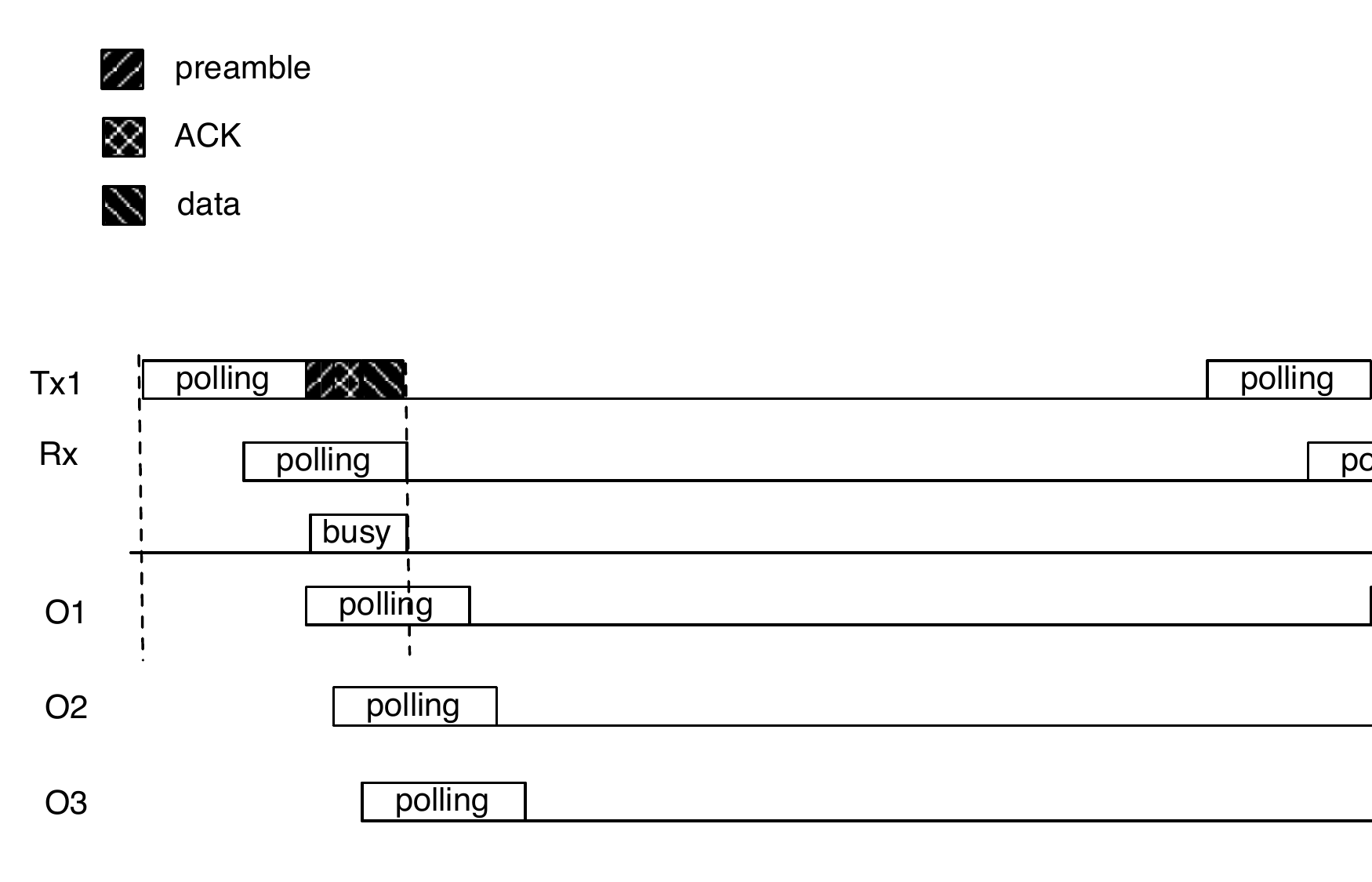}
\end{center}
\caption{Global buffer size A=1. Overhearing situations for Cases 2, 3 and 4. X-MAC protocol} 
\label{fig:xmacA1case234}
\end{figure}
\vspace{1cm}
\begin{itemize}
\item Case 2
If the ovehearer wakes up during a preamble transmission, it will ovehear the following ACK and afterwards go back to sleep.
The probability for the overhearer to wakeup during a preamble is $p_a = t_p^X/t_f$.
\begin{equation}
	E_{Case_2,o}^X = \frac{t_p^X}{2} \cdot P_l + t_a^X \cdot P_r+  (t_f - \frac{t_p^X}{2}- t_a^X) \cdot P_s
\end{equation}

\item Case 3:
If the ovehearer wakes up during a ACK transmission, it will ovehear the following data message and afterwards go back to sleep.
The probability for the overhearer to wakeup during an ACK is $p_b = t_a^X/t_f$.
\begin{equation}
	E_{Case_3,o}^X =\frac{t_a^X}{2} \cdot P_l + t_d \cdot P_r+  (t_f - \frac{t_a^X}{2}- t_d) \cdot P_s
\end{equation}
\item Case 4:
The ovehearer will either wakes up during data transmission or after the end of it. 
In both cases when the sender wakes up and senses the channel, it will sense it as free because the sender was sleeping when data packet transmission begun. 
Therefore the overhearer  performs an entire polling and go back to sleep.
The probability for this event to happen is $1-p_a-p_b$.
\begin{equation}
	E_{Case_4,o}^X = t_l \cdot P_l+  (t_f - t_l) \cdot P_s
\end{equation}
\end{itemize}
\item Case 5
\vspace{1cm}
\begin{figure}[ht!]
\begin{center}
\includegraphics[scale=.5]{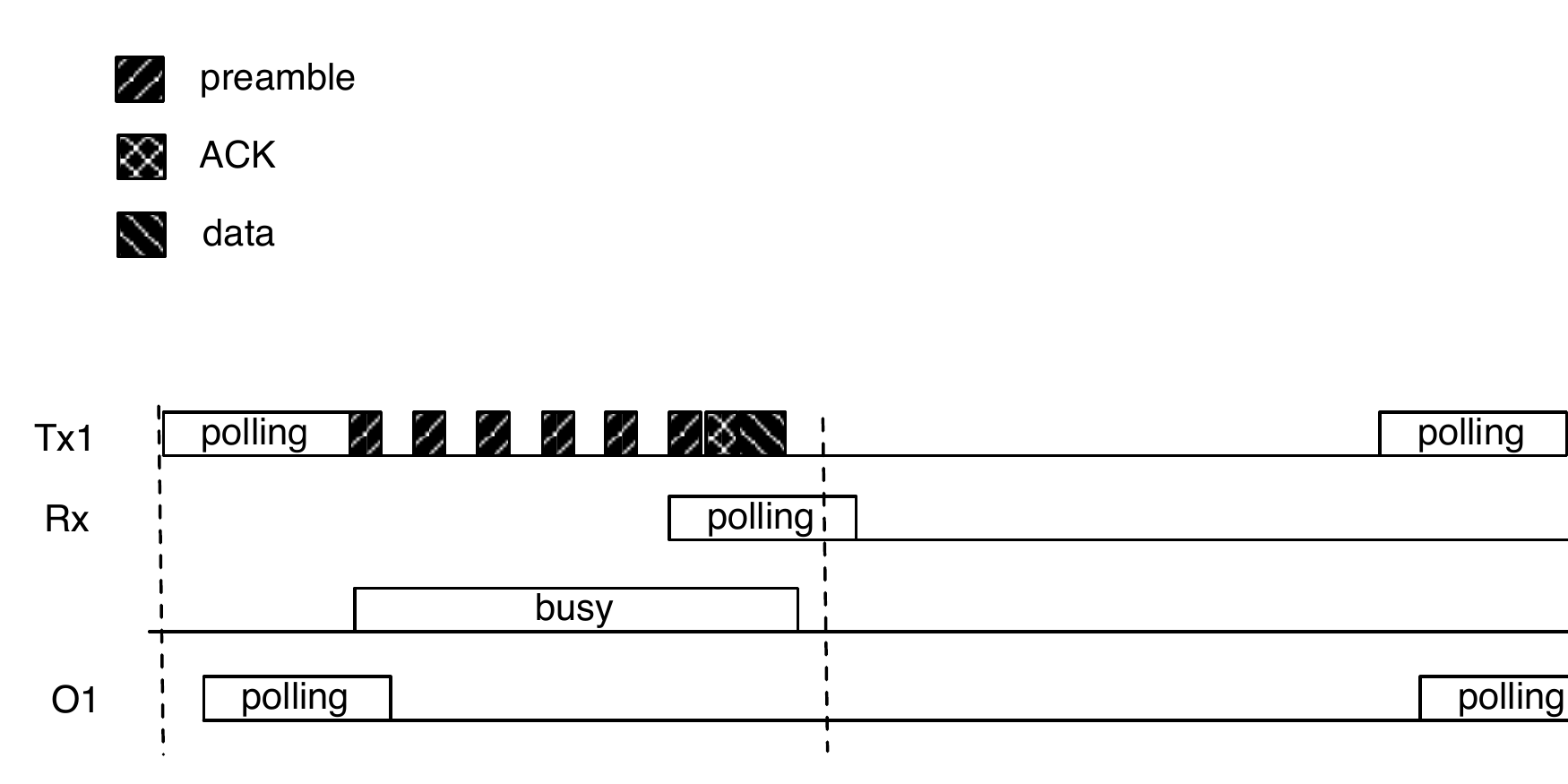}
\end{center}
\caption{Global buffer size A=1. Overhearing situations for Case 5. X-MAC protocol} 
\label{fig:xmacA1case5}
\end{figure}
\vspace{1cm}
Similarly to Case 1, if the overhearer is quasi-synchronized with the transmitter it will overhear the first preamble even if the receiver is still sleeping. 
The energy cost is:
\begin{equation}
	E_{Case_5,o}^X  = E_{Case_1,o}^X 
\end{equation}
\item Cases 6,7,8: If neither the receiver nor the overhearer are synchronized with the sender, it may happen that the receiver wakes up before the overhearer.	
Therefore, similarly to cases 2,3 and 4 we have different situations.
\vspace{1cm}
\begin{figure}[ht!]
\begin{center}
\includegraphics[scale=.5]{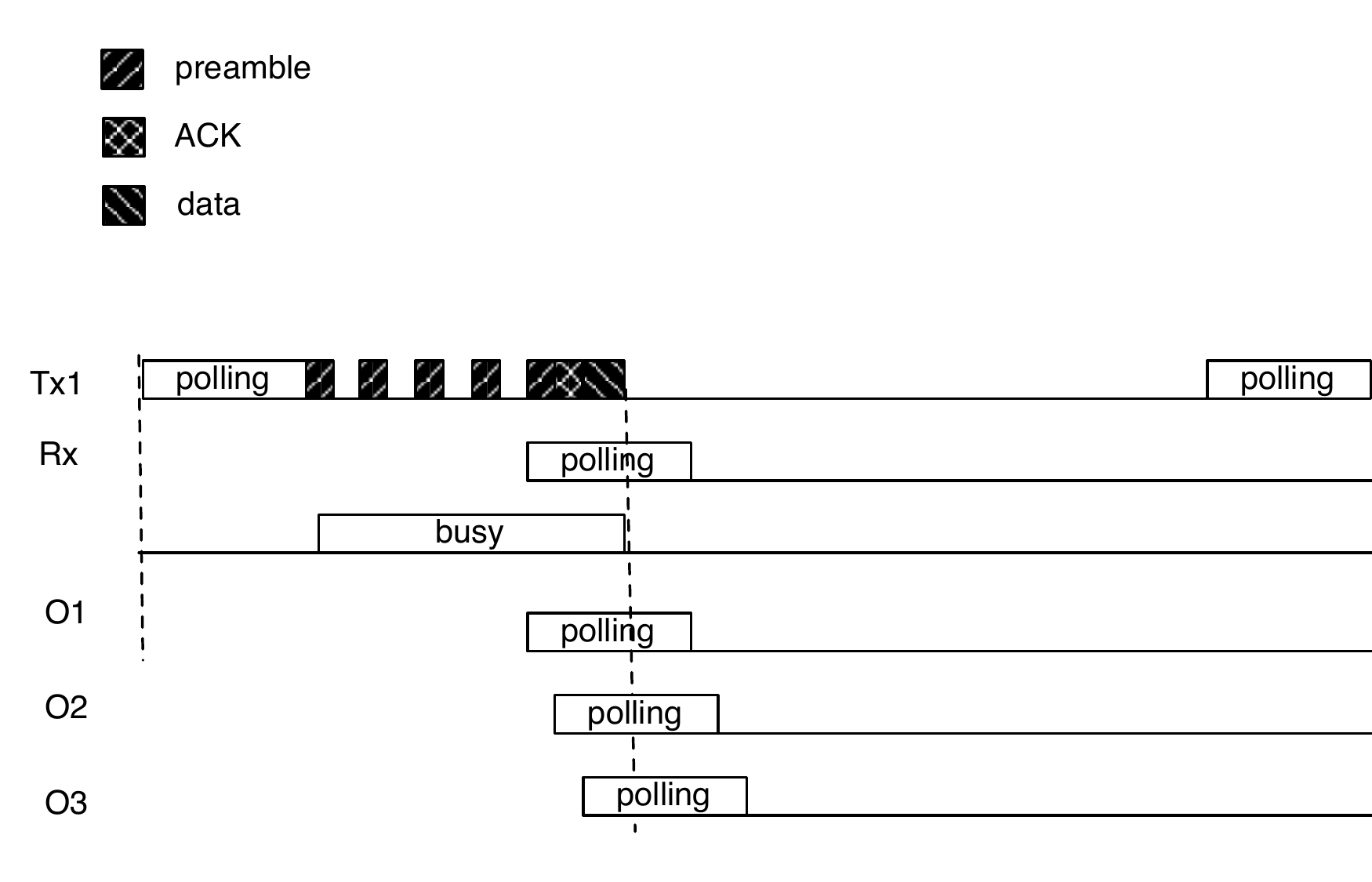}
\end{center}
\caption{Global buffer size A=1. Overhearing situations for Cases 6, 7 and 8. X-MAC protocol} 
\label{fig:xmacA1case678}
\end{figure}
\vspace{1cm}
Cases 6,7,8 are respectively similar to 2,3 and 4:
\begin{equation}
E_{Case_6,o}^X = E_{Case_2,o}^X 
\end{equation}
\begin{equation}
E_{Case_7,o}^X = E_{Case_3,o}^X  
\end{equation}
\begin{equation}
E_{Case_8,o}^X  = E_{Case_4,o}^X 
\end{equation}
\item Case 9: If the overhearer wakes up before the intended receiver, it will receive a preamble and go back to sleep.
The cost in this case is:
\begin{equation}
	E_{Case_9,o}^X = t_p^X \cdot P_r + \frac{t_p^X + t_a^X}{2} \cdot P_l+ (t_f -  \frac{t_p^X + t_a^X}{2} -t_p^X) \cdot P_s
\end{equation}
\end{itemize}

The overall energy cost is the sum of the costs of each case weighted by the probability of the case to happen

\begin{equation}
E_o^{X}(1) = N_o \cdot \sum\limits_{i=1}^{9} p_{Case_i}\cdot E_{Case_i,o}^X
\end{equation}


\textit{ }

\noindent\textbf{LA-MAC ($B=1$)}

In the present analysis we do not consider adaptive wakeup schedule of senders presented in section [REF PROTOCOL DESCRIPTION SECTION]. 
Therefore, wakeup schedules are assumed random. 
Even if this is a worst case for LA-MAC, it helps to better compare  it with other protocols. 
When sender wakes up, polls channel and send preambles as in X-MAC. 
However differently from X-MAC,  after \textit{early} ACK reception, the sender goes back to sleep and waits for Schedule message to be sent. 
When the intended receiver receives one preamble, it clears it and completes its polling period in order to detect other possible preambles to clear. 
Immediately after the end of polling period, the receiver processes requests and broadcasts the Schedule message. 
In LA-MAC,  overhearers go to sleep as soon as they receive any unicast message (preamble, ACK or data) as well as the Schedule.
Due to lack of synchronization, expected number of preambles per slot follows X-MAC with different size of preambles $t_p^L$ and ACK $t_a^L$.
When the sender wakes up,  it perform an entire channel polling before starting transmitting strobed preambles. 
When the receiver  wakes up, it polls the channel.
With probability  $p=t_l/t_f$ the sender and receiver are quasi-synchronized; so with probability $p$ the sender is still  polling the channel when the receiver wakes up.

When the sender wakes up, it polls the channel and starts sending preambles to \textit{wakeup} the receiver. 
With probability $p$, the first preamble that is sent will wake up the receiver, so the sender will immediately receive an \textit{early} ACK.
Otherwise, if nodes are not synchronized (probability $(1-p)$) the sender will wake up its destination in average after $\gamma^L$ preambles.
$E_t^{L}(1)$ is similar to the cost of X-MAC plus the cost of receiving the Schedule.
\begin{equation}
E_t^{L}(1) =  (1-p)\cdot  \gamma^L  \cdot  t_p^L  \cdot P_t  + p\cdot    t_p^L   \cdot P_t +  t_a^L \cdot  P_r  +  t_d \cdot  P_t + t_g \cdot  P_r 
\end{equation}
Cost of reception depends on the duration of preamble, ACK, data and Schedule messages.
\begin{equation}
E_r^{L}(1) =   (t_p^L +  t_d ) \cdot P_r +  (t_a^L + t_g  )\cdot P_t
\end{equation}
When the sender wakes up,  it performs a full polling period before the beginning of the strobed preambles. 
Moreover the degree of synchronization between sender and receiver (called \textit{active} nodes) also influences the consumption. 
If active nodes are not synchronized, the sender will poll the channel $(\gamma^L- 1)$ times in order to wait for \textit{early} ACK. 
Differently from X-MAC, the receiver will complete its polling period even if it clears one preamble, so  its radio will remain in polling state the duration of a full polling period less the time for preamble reception and ACK transmission.
\begin{equation}
E_l^{L}(1) =  ((t_l + (1-p) \cdot (\gamma^L- 1) \cdot t_a^L ) + (t_l - t_p^L - t_a^L  ))\cdot P_l
\end{equation}
When the active nodes are not transmitting, receiving or polling the channel they can sleep.
\begin{equation}
E_s^{L}(1) = ( 2\cdot t_f -(   t_l  + (1-p)\cdot  \gamma^L  \cdot t_p^L + p\cdot    t_p^L  +   t_a^L + (1-p) \cdot (\gamma^L- 1) \cdot t_a^L + t_d + t_g) - (t_l   + t_d + t_g )) \cdot P_s
\end{equation}

\begin{figure}[ht!]
\begin{center}
\includegraphics[scale=.5]{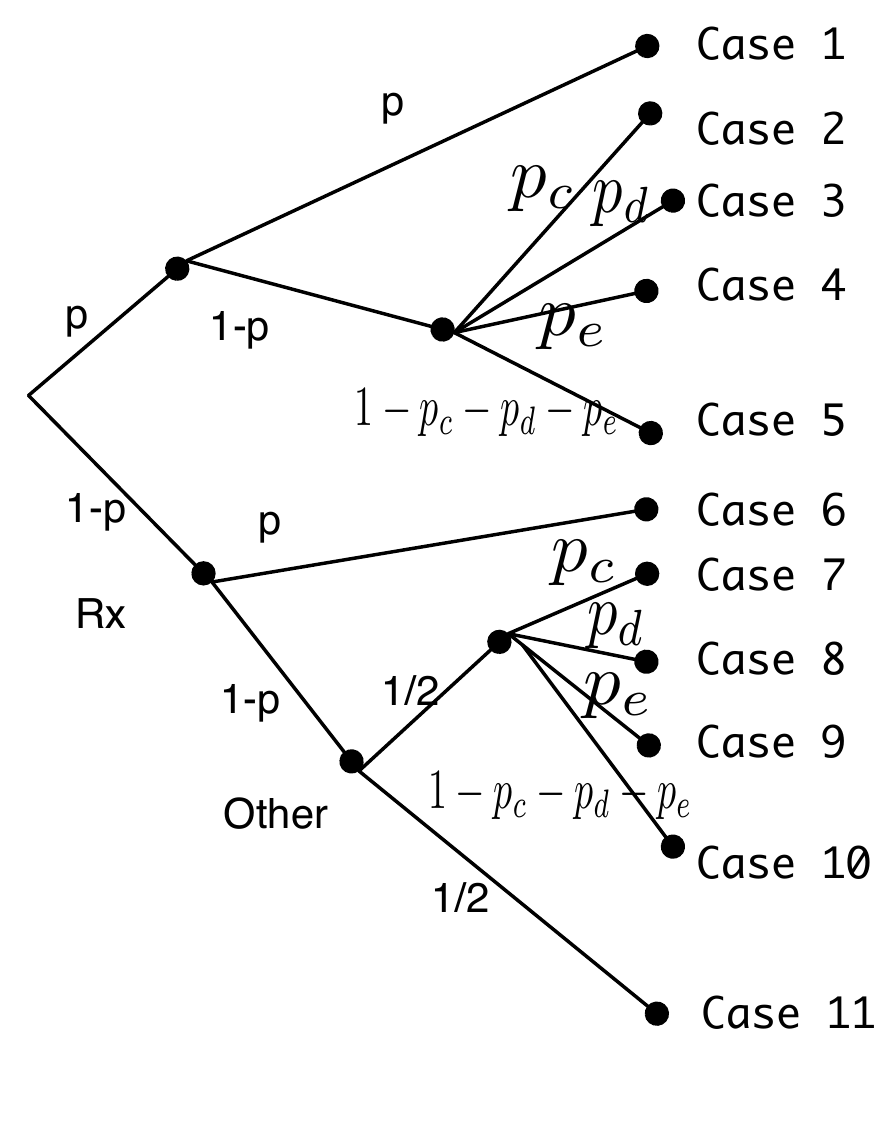}
\caption{Lamac. Tree of different wakeup cases.} 
\label{fig:probabilitiesOthersLamac}
\end{center}
\end{figure}

As in X-MAC as soon as overhearers receive some messages they go back to sleep. 
Therefore their energy consumption depends on the probability
that such nodes wake up while the channel is busy or not. 
All the possible combinations of wakeup schedules with relative probabilities are  shown in  figure~\ref{fig:probabilitiesOthersLamac} .

\vspace{1cm}
\vspace{1cm}
\begin{itemize}
\item Case 1 :  Sender, receiver and overhearer are quasi-synchronized. The  overhearer will sense a preamble that is not intended to it and goes back to sleep. 
Probability of this event is $p\cdot p $
\vspace{1cm}
\begin{figure}[ht!]
\begin{center}
\includegraphics[scale=.4]{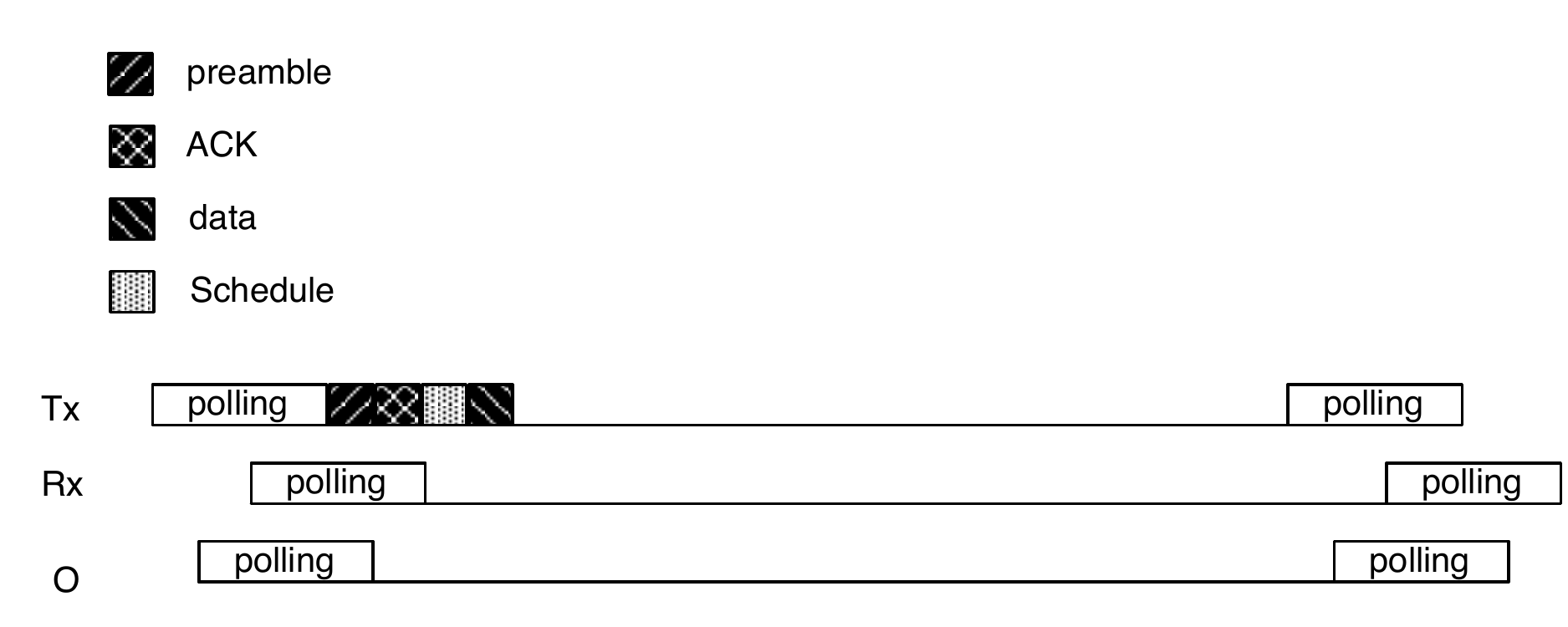}
\caption{Lamac. Possible wakeup instants of overhearers. Case 1} 
\end{center}
\end{figure}
\vspace{1cm}
\begin{equation}
	E_{Case_1,o}^L =\frac{t_l}{2} \cdot P_l+  t_p^L \cdot P_r +  (t_f - \frac{t_l}{2} -t_p^L) \cdot P_s
\end{equation}
\item Case 2, 3, 4, 5: the receiver is synchronized with Sender. Nevertheless, the  overhearer is not synchronized with the sender. 
When the overhearar  wakes up, it can receive different messages (preamble, ACK, Schedule or data) as well as clear channel.
\vspace{1cm}
\begin{figure}[ht!]
\begin{center}
\includegraphics[scale=.4]{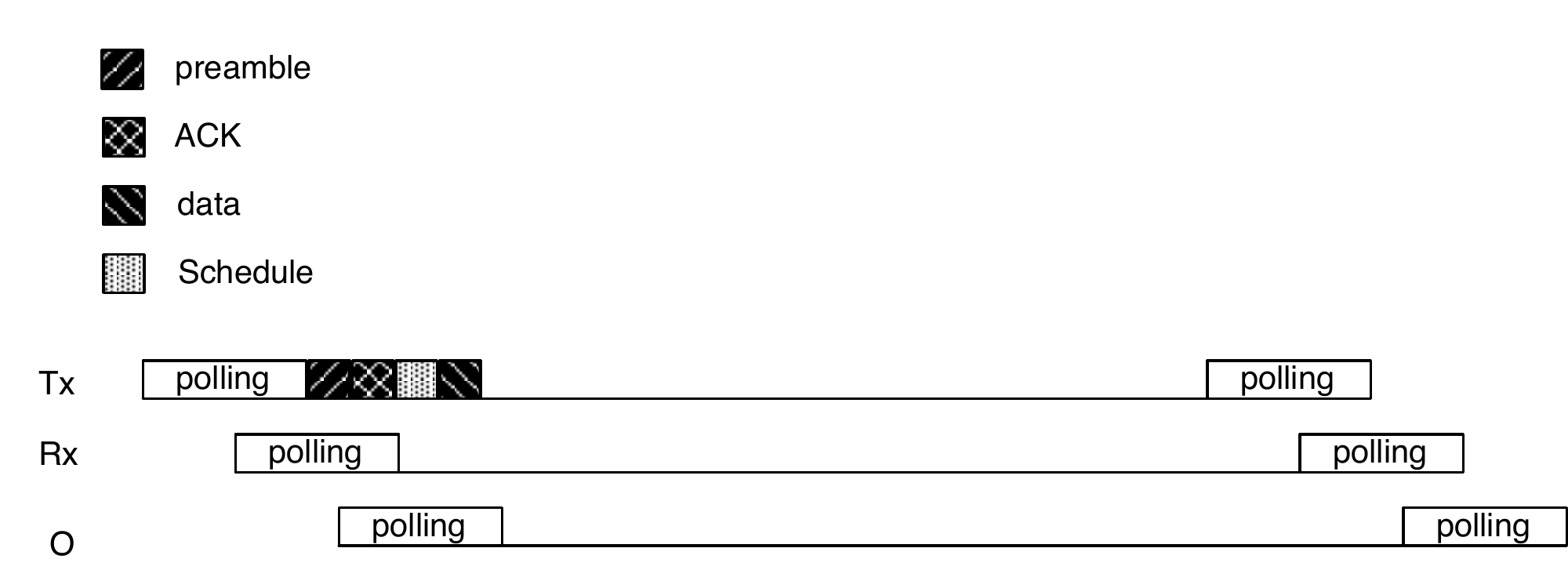}
\caption{Lamac. Possible wakeup instants of overhearers. Case 2}
\end{center}
\end{figure}
\vspace{1cm}
\begin{itemize}
\item Case 2
If the ovehearer wakes up during a preamble transmission, it will receive in average half of the preamble and ovehear  the following ACK. Afterwards it will go back to sleep.
Probability of this event is $p\cdot (1-p) \cdot p_c$, where $p_c = t_p^L/t_f$ represents the event that wakeup instant  of the overhearer is slightly after the end of polling of the sender. 
\begin{equation}
	E_{Case_2,o}^L = \frac{t_p^L}{2} \cdot P_l + t_a^L \cdot P_r  + (t_f - \frac{t_p^L}{2}- t_a^L) \cdot P_s
\end{equation}

\item Case 3:
If the ovehearer wakes up during an ACK transmission, it will sense a silent period and ovehear the following schedule message. Afterwards it goes back to sleep.
Probability of this event is $p\cdot (1-p) \cdot p_d$, where $p_d = t_a^L/t_f$ includes the event that wakeup instant  of the overhearer happens at least after the transmission of a preamble.  $p_d$ Neglects the time that elapses between the end of the ACK and the end of channel polling of the receiver. In other words,  $p_d$ supposes that schedule message is sent immediately after the transmission of ACK.
\begin{equation}
	E_{Case_3,o}^L =  \frac{t_a^L}{2} \cdot P_l +  t_g \cdot P_r+  (t_f - \frac{t_a^L}{2}- t_g) \cdot P_s
\end{equation}
\item Case 4:
If the overhearer wakes up during the transmission of the Schedule, it will hear the following data and then go to sleep. 
Probability of this event is $p\cdot (1-p) \cdot p_e$, where $p_e = t_g/t_f$ assumes that the wakeup instant  of the overhearer happens in average during the middle of schedule transmission. 
\begin{equation}
	E_{Case_4,o}^L =\frac{t_g}{2}\cdot P_l+ t_d \cdot P_r +   (t_f -\frac{t_g}{2} - t_d) \cdot P_s
\end{equation}
\item Case 5
The ovehearer will either wakes up during data transmission or will sense a free channel  because both sender and receiver  are already sleeping.  
Therefore the overhearer performs an entire polling and goes back to sleep.
Probability of this event is $p\cdot (1-p) \cdot (1-p_c-p_d-p_e)$. 
\begin{equation}
	E_{Case_5,o}^L = t_l \cdot P_l+  (t_f - t_l) \cdot P_s
\end{equation}
\end{itemize}

\item Case 6:
Similarly to Case 1, if the overhearer is quasi-synchronized with the sender with probability $(1-p)\cdot p$, the energy cost is:
\begin{equation}
	E_{Case_6,o}^L  = \frac{t_l}{2} \cdot P_l+  t_p^L \cdot P_r +  (t_f - \frac{t_l}{2} -t_p^L) \cdot P_s
\end{equation}
\item Cases 7,8,9,10: If neither the receiver nor the ovehearer are synchronized with sender, it may happen that the receiver wakes up before the overhearer.
We distinguish the situations of quasi-synchronization of the couple overhearer-preambles and lack of synchronization. 
\vspace{1cm}
\begin{figure}[ht!]
\begin{center}
\includegraphics[scale=.4]{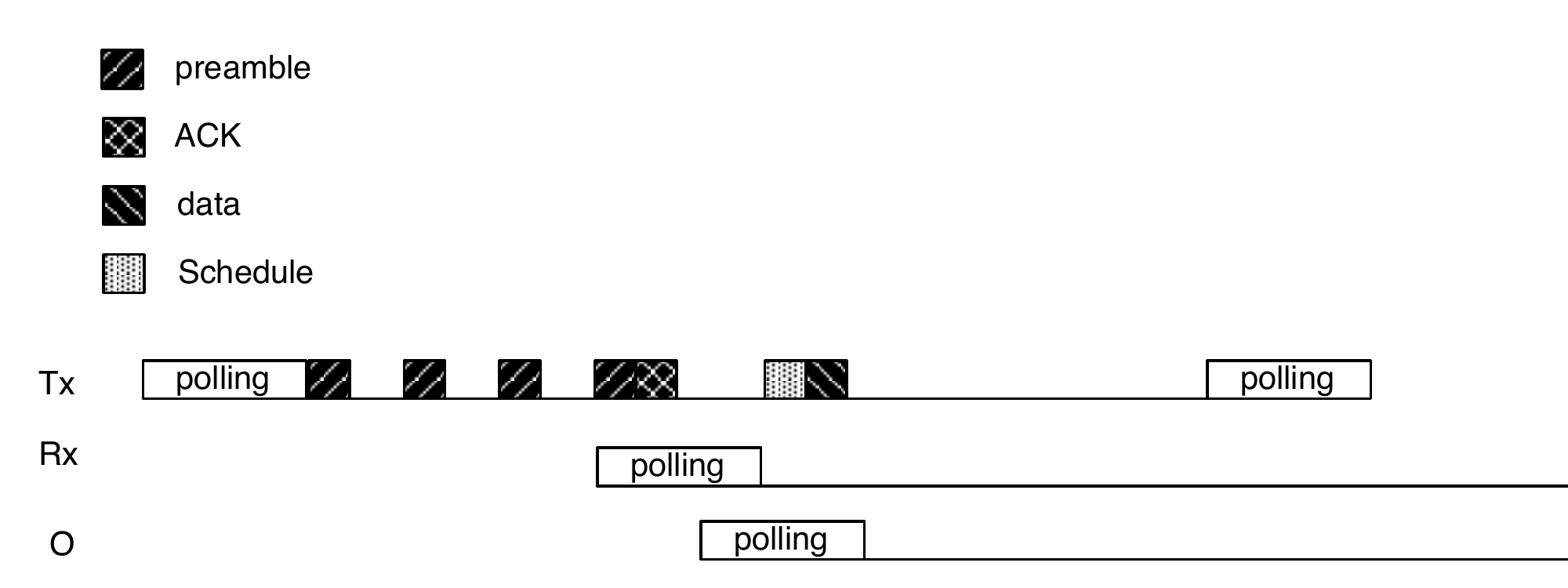}
\caption{Lamac. Possible wakeup instants of overhearers. Cases 7,8,9, 10.}
\end{center}
\end{figure}
\vspace{1cm}

In cases 7 and 8, overhearer is quasi-synchronized with the receiver.
\begin{itemize}
\item Case 7:
There is a probability to overhear a preamble. 
Such a probability is $(1-p) \cdot  (1-p) \cdot 1/2 \cdot p_c$. 
Consumption of this case is the same of Case 2.
\begin{equation}
E_{Case_7,o}^L = E_{Case_2,o}^L 
\end{equation}
\item Case 8:
There is a probability to overhear an ACK. 
Such a probability is $(1-p) \cdot  (1-p) \cdot 1/2 \cdot p_d$. 
Consumption of this case is the same of Case 2.
\begin{equation}
E_{Case_8,o}^L =E_{Case_3,o}^L 
\end{equation}
\end{itemize}
If the overhearer and the receiver are not synchronized:
\begin{itemize}
\item Case 9:
There is a probability to overhear a Schedule.
Such a probability is $(1-p) \cdot  (1-p) \cdot 1/2 \cdot p_e$. 
Consumption of this case is the same of Case 4.
\begin{equation}
E_{Case_9,o}^L = E_{Case_4,o}^L 
\end{equation}
\item Case 10:
There is a probability to overhear a data message. 
Such a probability is $(1-p) \cdot  (1-p) \cdot 1/2 \cdot (1-p_c-p_d-p_e)$. 
Consumption of this case is the same of Case 5.
\begin{equation}
E_{Case_{10},o}^L =E_{Case_5,o}^L 
\end{equation}
\end{itemize}

\item Case 11: Otherwise, if the overhearer wakes up before the intended receiver, it will receive one preamble (whichever preamble amongst $\gamma^L$) and go back to sleep.
The cost in this case is:
\begin{equation}
	E_{Case_{11},o}^L = \frac{t_p^L + t_a^L}{2} \cdot P_l + t_p^L \cdot P_r + (t_f -  \frac{t_p^L + t_a^L}{2} -t_p^L) \cdot P_s
\end{equation}
\end{itemize}

The overall energy cost is the sum of the costs of each case weighted by the probability of the case to happen

\begin{equation}
E_o^{L}(1) = N_o \cdot \sum\limits_{i=1}^{11} p_{Case_i}\cdot E_{Case_i,o}^L
\end{equation}

\subsection{Global buffer contains two messages ($B=2$)}
If \textit{A=2}, there can be either one sender with two messages to deliver, or two senders with each only one message. The others devices may overhear some channel activity.
The number of overhearers will be $N_o=N-1$ if there is just one sender, $N_o=N-2$ otherwise.
The probability that two messages are in different buffers is equal to $(N-1)/N$.


\textit{ }

\noindent\textbf{B-MAC ($B=2$)}

The overall power consumption for transmission and reception when $A\geq1$ is linear with the global number of packets in buffer, independently on how packets are distributed in the different buffers,\textit{i.e.,} independently of the number of senders. 
In fact, due to the long preamble to send ($t_p^b=t_f$), there can be only one sender per frame. 
Thus, we have the following relation: $E^B(A) = A\cdot E^B(1) = A\cdot (E_t^B(1) + E_r^B(1) + E_l^B(1) +  E_s^B(1) + E_o^B(1))$.

Such a relation depicts the limitations of B-MAC protocol, since high-loaded traffic can hardly been addressed.


\textit{ }

\noindent\textbf{X-MAC ($B=2$)}

After the reception of the first data message, the receiver remains in polling state for an extra back-off time $t_b$ during which it can receive a second message.
The energy consumed for the transmission of the first packet is the same as the energy $E_t^{X}(1)$ defined in the previous subsection; then the cost of the transmission for the second message must be added.

Differently from B-MAC, the distribution of messages in the buffers impacts X-MAC protocol behaviour.
With probability $1/N$ both packets are in the same buffer; otherwise two different senders are implicated, so we need to study how wakeup instants of the active agents are scheduled with respect to each others.
Wakeup instant of different agents are all independent. 
We assume that the frame begins at the wakeup instant of the first transmitter; scenarios that may happen are illstrated on Figure~\ref{fig:probabilitiesXmac} with their happening probability:
\vspace{1cm} 

\begin{figure}[ht!]
\begin{center}
\includegraphics[scale=.5]{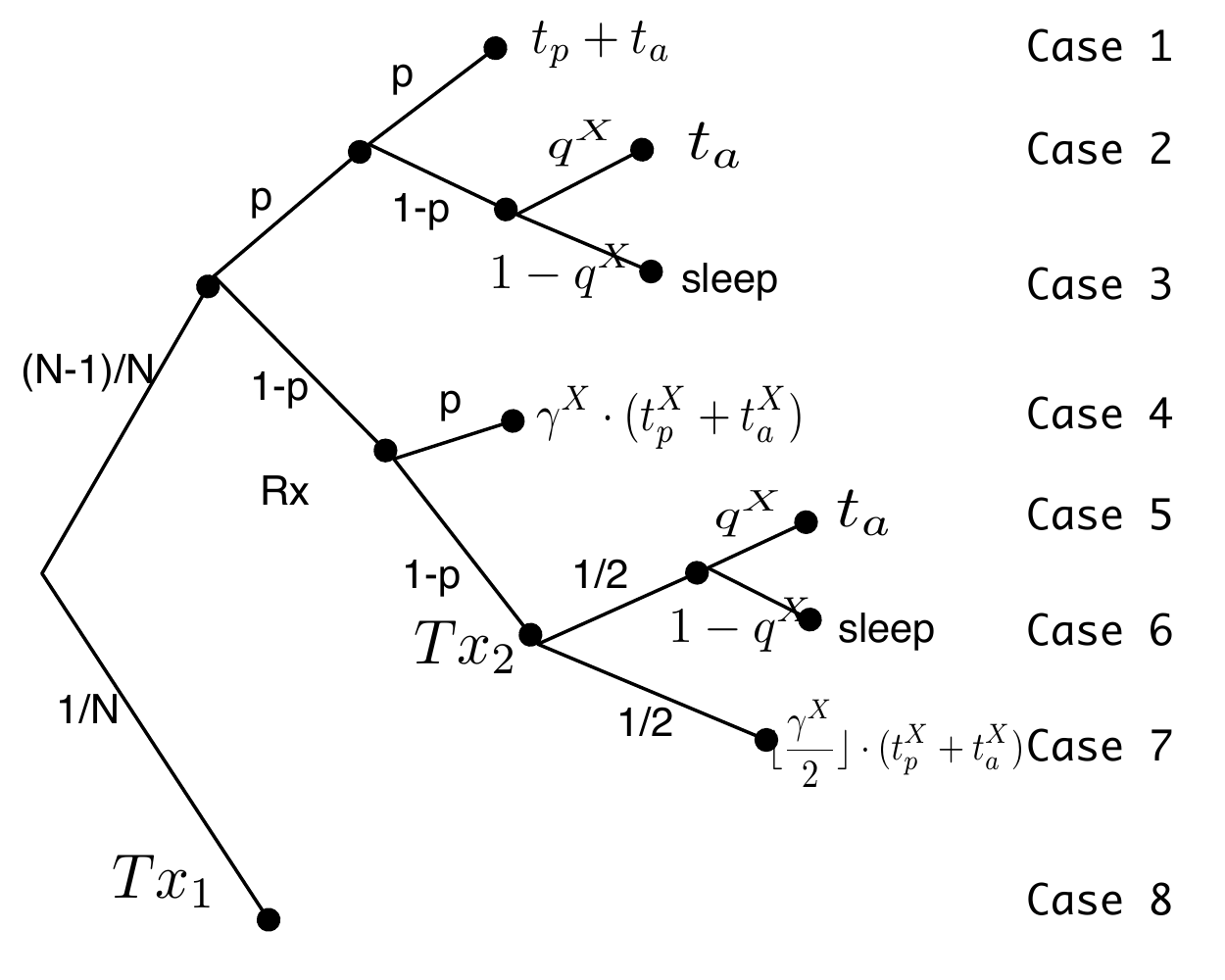}
\end{center}
\caption{X-MAC protocol: Probability tree of wakeup combinations with global buffer size A=2. There are one sender and  one or two transmitters.}
\label{fig:probabilitiesXmac}
\end{figure}
\vspace{1cm}

\begin{itemize}
\item Case 1: All three agents are quasi-synchronized.
The very first preamble sent by the first transmitter is cleared by the receiver who sends an ACK; the second transmitter hears both the preamble and the ACK. 
Probability of this scenario is $p_{Case_1}=(N-1)/N\cdot p\cdot p$.
\vspace{0.5cm} 

\begin{figure}[h!]
\begin{center}
\includegraphics[scale=.5]{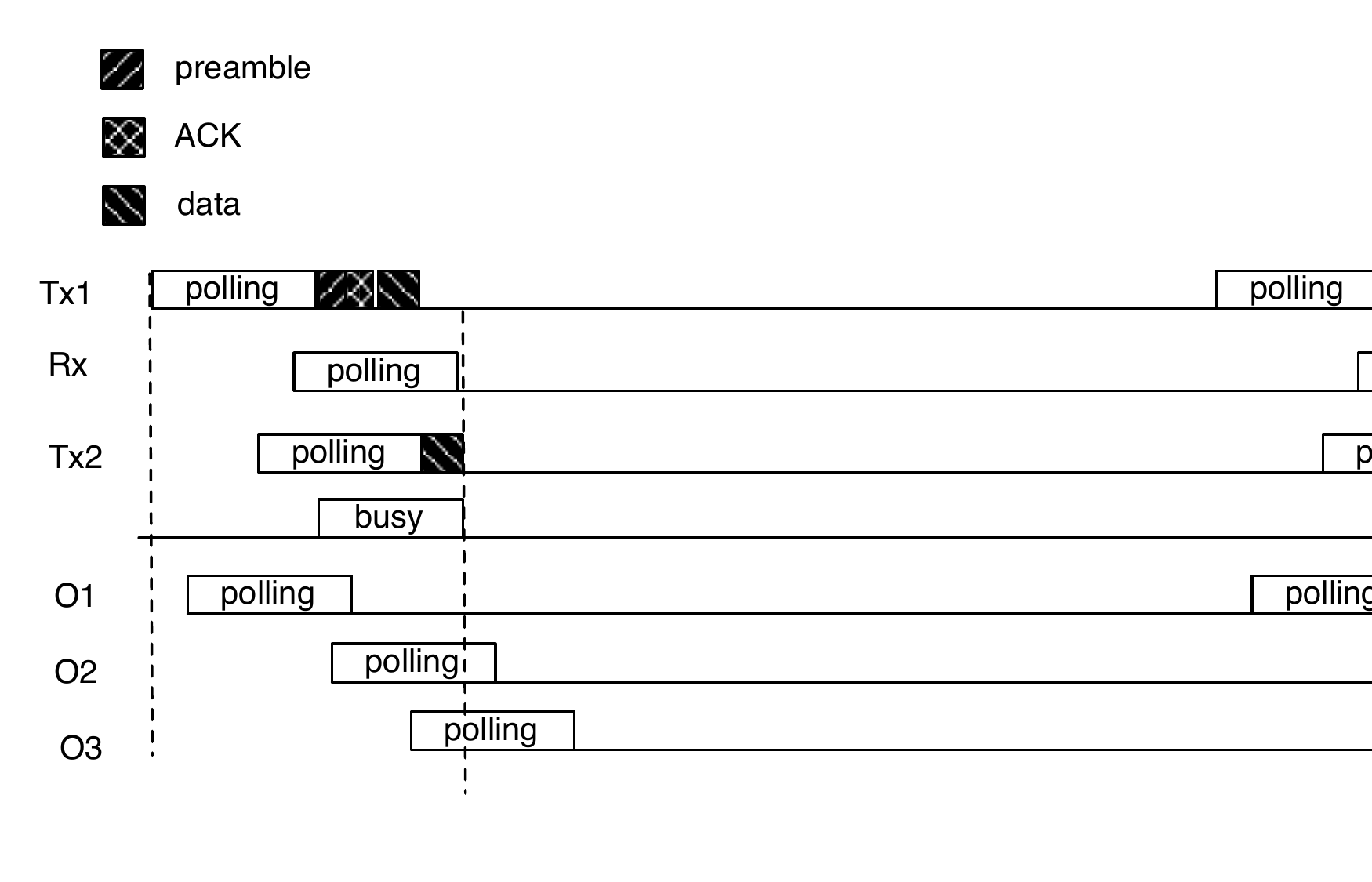}
\end{center}
\caption{X-MAC protocol, global buffer size A=2: Overhearing situations for Case 1.} 
\label{fig:xmacA2case1}
\end{figure}
\vspace{1cm} 

\begin{equation}
E_{Case_1,t}^X(2) = t_p^X\cdot P_t + t_a^X\cdot P_r + (t_p^X+t_a^X)\cdot P_r + 2\cdot t_d\cdot P_t  
\end{equation}

\begin{equation}
E_{Case_1,r}^X(2) = (t_p^X+2\cdot t_d)\cdot P_r+t_a^X\cdot P_t
\end{equation}

\begin{equation}
E_{Case_1,l}^X(2) = (t_l+\frac{t_l}{2}+\frac{t_l}{2})\cdot P_l  
\end{equation}

\begin{equation}
E_{Case_1,s}^X(2) = (3\cdot t_f - (t_l+t_p^X+t_a^X+t_d) - (\frac{t_l}{2}+t_p^X+t_a^X+t_d) - (\frac{t_l}{2}+t_p^X+t_a^X+2\cdot t_d))\cdot P_s 
\end{equation}

Depending on wakeup instants of overhearers several situations may happen.
If the overhearer is quasi-synchronized with one of the three active agents (receiver or one of the two senders), then it will sense a busy channel (cf. figure~\ref{fig:xmacA2case1}).
We assume that an overhearer polls the channel for some time and then overhears a message that can be a preamble, an ACK or a data.
For simplicity, we assume the overhearer polls the channel during in average a half polling frame and then overhears a data (the largest message that can be overheard). 
Probability to wakeup during a busy period is $p_{case1,A=2}^X=(t_p^X+t_a^X+2\cdot t_d)/t_f$.
Otherwise, the overhearer wakes up while channel is free; it polls the channel and then goes back to sleep.

\begin{equation}
E_{Case_1,o}^X(2) = N_o\cdot(p_{case1,A=2}^X\cdot(\frac{t_l}{2}\cdot P_l+t_d\cdot P_r+(t_f-\frac{t_l}{2}-t_d)\cdot P_s) + (1-p_{case1,A=2}^X)\cdot(t_l\cdot P_l+(t_f-t_l)\cdot P_s))
\end{equation}

\item Case 2: First sender and receiver are quasi-synchronized, contrary to second sender (cf. figure~\ref{fig:xmacA2case2}). 
The only possibility for the second sender to send data in the current frame is to manage to catch the ACK of the receiver during its polling period.
This event happens with probability $q^X=(t_l-t_a^X)/t_f$.
Probability of this scenario is $p_{Case_2}=(N-1)/N\cdot p\cdot(1-p)\cdot q^X$.
\vspace{0.5cm}

\begin{figure}[h!]
\begin{center}
\includegraphics[scale=.5]{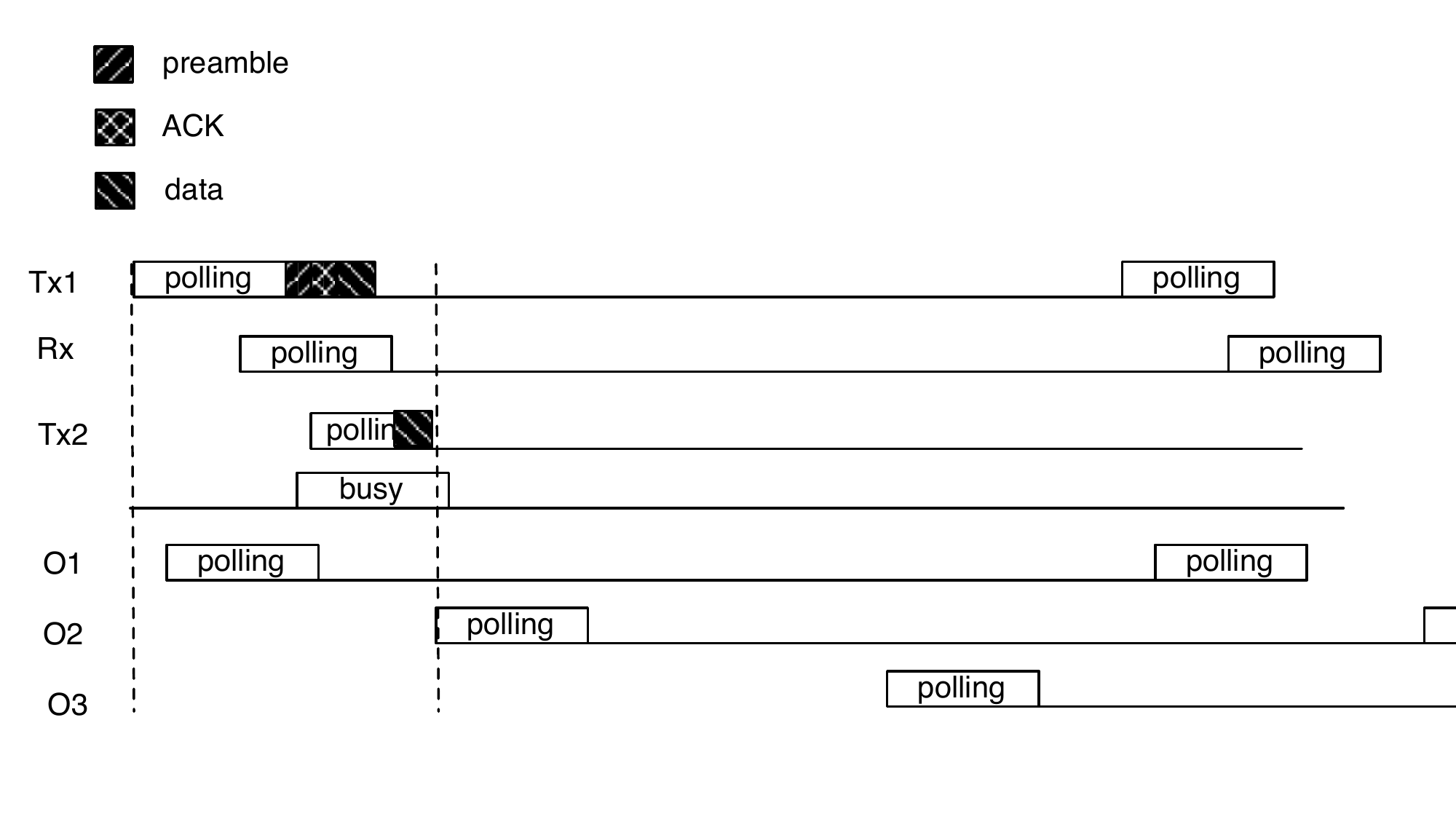}
\end{center}
\caption{X-MAC protocol, global buffer size A=2: Overhearing situations for Case 2.} 
\label{fig:xmacA2case2}
\end{figure}
\vspace{1cm}

Energy consumption of this second scenario is quite the same as the one of Case 1 but event probability is different.
Since the second sender is not quasi-synchronised, it cannot hear the full preamble sent by the first sender and has a shorter polling period.
\begin{equation}
E_{Case_2,t}^X(2) = E_{Case_1,t}^X(2)-t_p^X\cdot P_r
\end{equation}
\begin{equation}
E_{Case_2,r}^X(2) = E_{Case_1,r}^X(2)
\end{equation}
\begin{equation}
E_{Case_2,l}^X(2) = E_{Case_1,l}^X(2)-\frac{t_l-t_p^X}{2}\cdot P_l
\end{equation}
\begin{equation}
E_{Case_2,s}^X(2) = E_{Case_1,s}^X(2) +  \frac{t_l +  t_p^X }{2}\cdot P_l
\end{equation}
We assume that the probability of busy channel is the same as the previous scenario. 
So, overhearing consumption is unchanged.
\begin{equation}
E_{Case_2,o}^X(2) = E_{Case_1,o}^X(2)
\end{equation}

\item Case 3: With probability $1-q^X$, the second sender wakes up too late and cannot catch the ACK. 
In this case, it goes back to sleep and it will transmit its data during the next frame.
So, energy cost is the sum of the transmission cost for the first packet in the current frame and for the second packet in the following frame.
This second frame is the same as $E^X(1)$.
This scenario happens with probabiliy $p_{Case_3}=(N-1)/N\cdot p\cdot(1-p)\cdot(1-q^X)$.

\begin{equation}
E_{Case_3,t}^X(2) = t_p^X\cdot P_t+t_a^X\cdot P_r+t_d\cdot P_t+ E_t^{X}(1) 
\end{equation}
\begin{equation}
E_{Case_3,r}^X(2) = t_p^X\cdot P_r+t_a^X\cdot P_t+t_d\cdot P_r+E_r^{X}(1) 
\end{equation}
\begin{equation}
E_{Case_3,l}^X(2) = (t_l+t_l+\frac{t_l}{2})\cdot P_l+E_l^{X}(1)
\end{equation}
\begin{equation}
E_{Case_3,s}^X(2) = (3\cdot t_f - (t_l+t_p^X+t_a^X+t_d) - t_l - (\frac{t_l}{2}+t_p^X+t_a^X+t_d))\cdot P_s+E_s^{X}(1)
\end{equation}

In the second frame, the first sender has nothing to send any more and can be count as an overhearer.
Then the number of overhearers should be updated between two frames but energy cost per overhearer is unchanged in comparison to the one for a unique message in global buffer (A+1):
\begin{equation}
E_{Case_3,o}^X(2) = (N_o+(N_o+1))\cdot\frac{E_o^{X}(1)}{N_o+1}
\end{equation}

\item Case 4: First and second senders are quasi-synchronized but the receiver wakes up later.
In this scenario, the first sender sends a strobed preamble until the receiver wakes up and sends an ACK; the second sender hears the whole strobed preamble and then sends its data during the back-off time.
Between short preambles, senders poll channel waiting for an ACK from receiver.
Probability of this scenario is $p_{Case_4}=(N-1)/N\cdot(1-p)\cdot p$.
\vspace{0.5cm}

\begin{figure}[h!]
\begin{center}
\includegraphics[scale=.5]{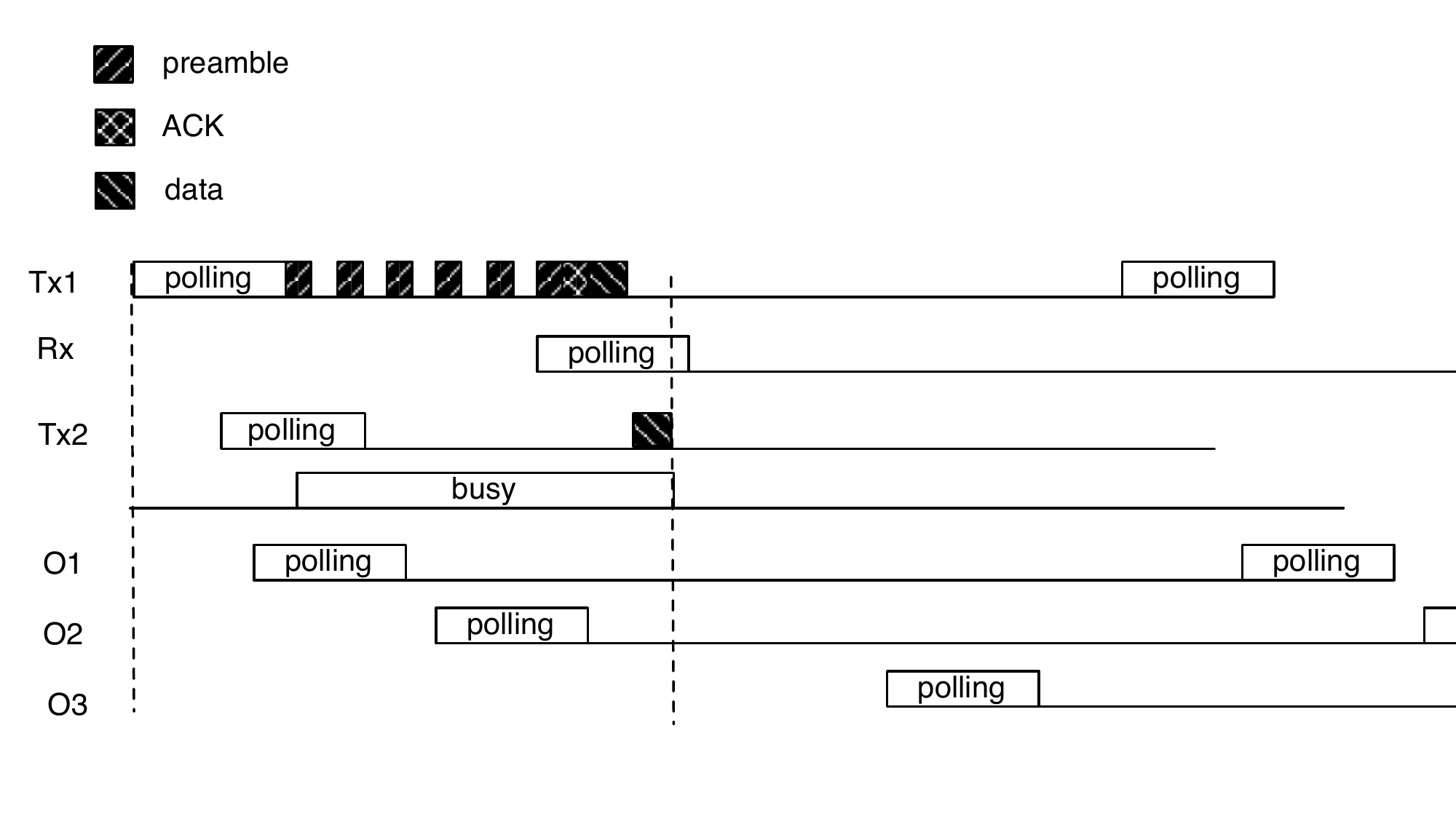}
\end{center}
\caption{X-MAC protocol, global buffer size A=2: Overhearing situations for Case 4.} 
\label{fig:xmacA2case4}
\end{figure}
\vspace{1cm}

\begin{equation}
E_{Case_4,t}^X(2) = \gamma^X\cdot t_p^X\cdot(P_t+P_r)+2\cdot t_a^X\cdot P_r +2\cdot t_d\cdot P_t
\end{equation}
\begin{equation}
E_{Case_4,r}^X(2) = (t_p^X+2\cdot t_d)\cdot P_r+t_a^X\cdot P_t
\end{equation} 
\begin{equation}
E_{Case_4,l}^X(2) = (t_l+\frac{t_l}{2}+2\cdot(\gamma^X-1)\cdot t_a^X+\frac{t_p^X+t_a^X}{2})\cdot P_l
\end{equation}
\begin{equation}
E_{Case_4,s}^X(2) = (3\cdot t_f - (t_l+\gamma^X\cdot(t_p^X+t_a^X)+t_d) - (\frac{t_l}{2}+\gamma^X\cdot(t_p^X+t_a^X)+t_d) - (\frac{t_p^X+t_a^X}{2}+t_p^X+t_a^X+2\cdot t_d))\cdot P_s
\end{equation}

When receiver wakes up later than both senders, the probability that an overhearer wakes up during a transmission of a preamble is higher than with previous scenarios. 
If this happens, the overhearer performs a very short polling, overhears a message (most probably a preamble) and then goes back to sleep.
For simplicity we assume that the overhearer will perform half of $(t_p^X + t_a^X)$ of polling and than overhears an entire preamble.
Probability of busy channel is thus $ p_{case4,A=2}^X =(\gamma^X \cdot(t_p^X +t_a^X)+2\cdot t_d)/t_f$.

\begin{equation}
E_{Case_4,o}^X(2) = N_o\cdot(p_{case4,A=2}^X\cdot(\frac{t_p^X+t_a^X}{2}\cdot P_l+t_p^X\cdot P_r+(t_f-\frac{t_p^X+t_a^X}{2}-t_p^X )\cdot P_s) + (1-p_{case4,A=2}^X)\cdot(t_l\cdot P_l+(t_f-t_l)\cdot P_s))
\end{equation}

\item Cases 5, 6, 7: Second sender and receiver are not synchronized with first sender; the behaviour of the protocol depends on which device among the second sender and the receiver will wake up as first.   

\begin{itemize}
\item Case 5: Receiver wakes up as first.
Similarly to Case 2, the only possibility for the second transmitter to send data in the current frame is to catch the ACK of the receiver during its polling. 
This event happens with probability $q^X=(t_l-t_a^X)/t_f$.
However, there is also the possibility for $Tx_2$ to catch the preamble sent by $Tx_1$ that just precedes the overheard ACK. 
Such eventuality can happen with probability $u^X = \dfrac{t_p^X+t_a^X}{2\cdot t_p^X+t_a^X }$.
This scenario happens with probability $p_{Case_5}=(N-1)/N\cdot(1-p)\cdot(1-p)\cdot\frac{1}{2}\cdot q^X$.
\vspace{0.5cm}

\begin{figure}[h!]
\begin{center}
\includegraphics[scale=.5]{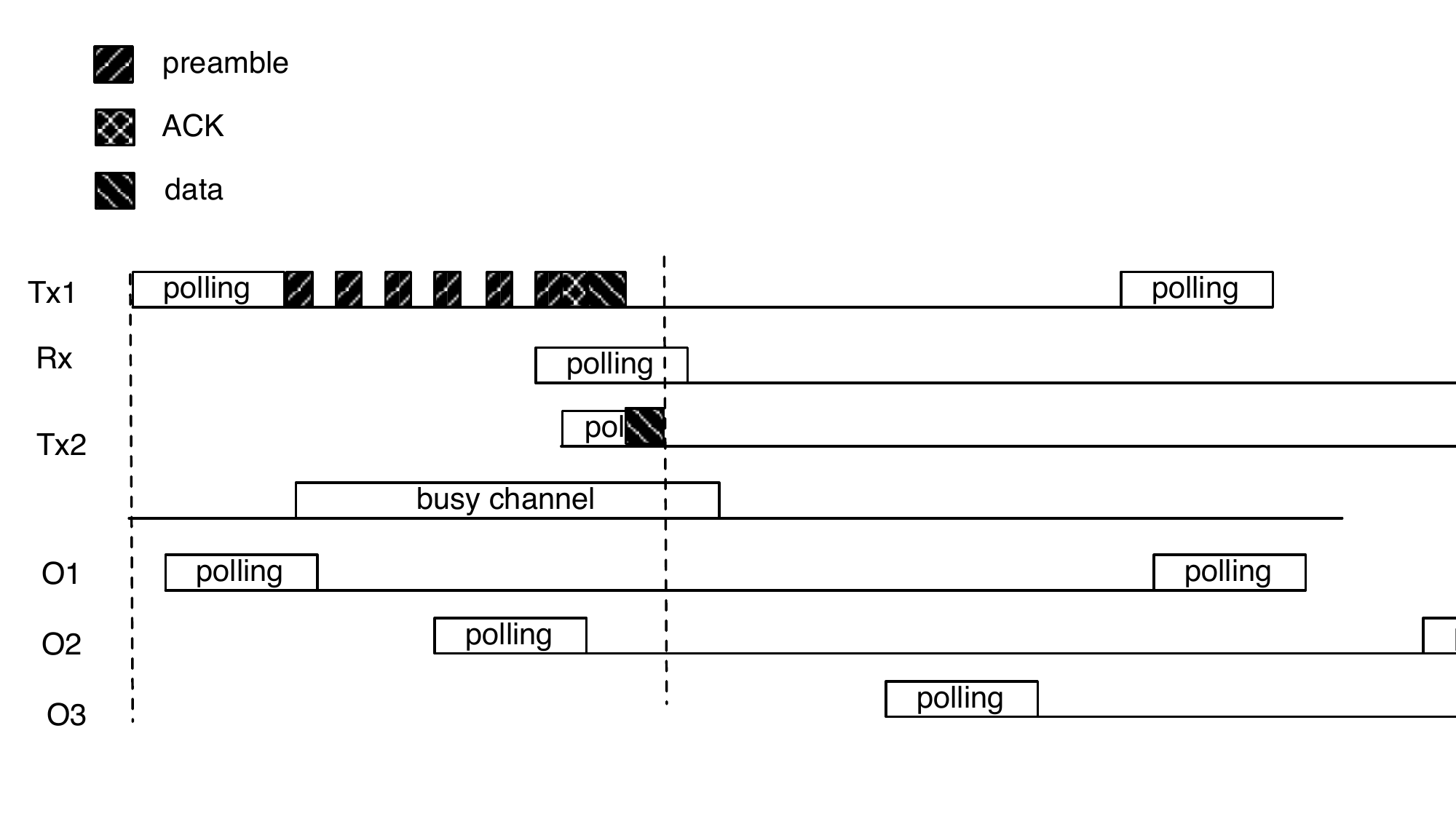}
\end{center}
\caption{X-MAC protocol, global buffer size A=2: Overhearing situations for Case 5.} 
\label{fig:xmacA2case5}
\end{figure}
\vspace{1cm}

\begin{equation}
E_{Case_5,t}^X(2) = (\gamma^X\cdot t_p^X+t_d)\cdot P_t+t_a^X\cdot P_r+(u^X\cdot t_p^X+t_a^X)\cdot P_r+t_d\cdot P_t
\end{equation}
\begin{equation}
E_{Case_5,r}^X(2) = (t_p^X+2\cdot t_d)\cdot P_r+t_a^X\cdot P_t
\end{equation}
\begin{equation}
E_{Case_5,l}^X(2) = (t_l+(\gamma^X-1)\cdot t_a^X + \frac{t_p^X+t_a^X}{2} + u^X\cdot\frac{t_p^X+t_a^X}{2}+(1-u^X)\cdot\frac{t_p^X}{2})\cdot P_l
\end{equation}
\begin{equation}
E_{Case_5,s}^X(2) = (3\cdot t_f - (t_l+\gamma^X\cdot(t_p^X+t_a^X)+t_d) - (u^X\cdot\frac{t_p^X+t_a^X}{2}+(1-u^X)\cdot\frac{t_p^X}{2}+u^X\cdot t_p^X+t_a^X+t_d) - (\frac{t_p^X+t_a^X}{2}+t_p^X+t_a^X+2\cdot t_d))\cdot P_s
\end{equation}

As in the previous case, the overhearer perceives a very busy channel because of the transmission of preambles; so when it wakes up it will perform half of $(t_p^X+t_a^X)$ in polling state before overhearing an entire preamble.
Probability of busy channel is $p_{case5}^X = p_{case4}^X$.

\begin{equation}
E_{Case_5,o}^X(2) =E_{Case_4,o}^X(2) 
\end{equation}

\item Case 6: Receiver wakes up as first.
Similarly to Case 3, with probability $1-q^X$, the second sender wakes up too late and cannot catch the ACK from the receiver.
Thus it goes back to sleep and will transmit its data during the next frame.
This scenario happens with probability $p_{Case_6}=(N-1)/N\cdot(1-p)\cdot(1-p)\cdot\frac{1}{2}\cdot(1-q^X)$.

\begin{equation}
E_{Case_6,t}^X(2) = \gamma^X\cdot t_p^X\cdot P_t+t_a^X\cdot P_r+t_d\cdot P_t+E_t^{X}(1)
\end{equation}
\begin{equation}
E_{Case_6,r}^X(2) = (t_p^X+t_d)\cdot P_r+t_a^X\cdot P_t+E_r^{X}(1)
\end{equation}
\begin{equation}
E_{Case_6,l}^X(2) = (t_l+(\gamma-1)\cdot t_a^X)\cdot P_l+t_l\cdot P_l+\frac{t_p^X+t_a^X}{2}\cdot P_l+E_l^{X}(1)
\end{equation}
\begin{equation}
E_{Case_3,s}^X(2) = (3\cdot t_f - (t_l+\gamma^X\cdot(t_p^X+t_a^X)+t_d) + t_l + (\frac{t_p^X+t_a^X}{2}+t_p^X+t_a^X+t_d))\cdot P_s+E_s^{X}(1)
\end{equation}
\begin{equation}
E_{Case_6,o}^X(2) = E_{Case_3,o}^X(2) = 2\cdot E_o^{X}(1)
\end{equation}

\item Case 7: Second transmitter wakes up as first, it hears a part of the strobed preamble until the receiver wakes up and sends its ACK. 
In average, when the second transmitter wakes up, it performs a short polling whose duration is the one between two successive short preambles: $\frac{t_p^X+t_a^X}{2}$.
After that, it hears an average number of $\lfloor\dfrac{\gamma^X}{2}\rfloor$ short preambles before the receiver wakes up and stops the strobed preamble by sending its ACK.
Probability of this scenario is $p_{Case_7}=(N-1)/N\cdot(1-p)\cdot(1-p)\cdot\frac{1}{2}$.

\begin{equation}
E_{Case_7,t}^X(2) = (\gamma^X\cdot t_p^X+t_d)\cdot P_t+t_a^X\cdot P_r+(\lfloor\frac{\gamma^X}{2}\rfloor\cdot t_p^X+t_a^X)\cdot P_r+t_d\cdot P_t
\end{equation}
\begin{equation}
E_{Case_7,r}^X(2) = (t_p^X+t_d)\cdot P_r+t_a^X\cdot P_t+t_d\cdot P_r
\end{equation}
\begin{equation}
E_{Case_7,l}^X(2) = (t_l+(\gamma^X-1)\cdot t_a^X)\cdot P_l+((\lfloor\frac{\gamma^X}{2}\rfloor-1)\cdot t_a^X+\frac{t_p^X+t_a^X}{2})\cdot P_l+\frac{t_p^X+t_a^X}{2}\cdot P_l
\end{equation}
\begin{equation}
E_{Case_7,s}^X(2) = (3\cdot t_f - (t_l+\gamma^X\cdot(t_p^X+t_a^X)+t_d) - (\frac{t_p^X+t_a^X}{2}+ \lfloor\frac{\gamma^X}{2}\rfloor\cdot(t_p^X+t_a^X)+t_d) - (\frac{t_p^X+t_a^X}{2}+t_p^X+t_a^X+2\cdot t_d))\cdot P_s
\end{equation}

From the overhearers point of view, this case is equivalent to Cases 4 and 5.
\begin{equation}
E_{Case_7,o}^X(2) = E_{Case_4,o}^X(2)
\end{equation}
\end{itemize}

\item Case 8: There is only one sender that sends two messages in a row during the extra back-off time.
This last scenario happens with a probability equal to $p_{Case_8}=\frac{1}{N}$.
\begin{equation}
E_{Case_8,t}^X(2) = E_t^{X}(1)+t_d\cdot P_t
\end{equation}
\begin{equation}
E_{Case_8,r}^X(2) = E_r^{X}(1)+t_d\cdot P_r
\end{equation}
\begin{equation}
E_{Case_8,l}^X(2) = E_l^{X}(1)-t_d\cdot P_l
\end{equation}
\begin{equation}
E_{Case_8,s}^X(2) = E_s^{X}(1)-t_d\cdot P_s
\end{equation}

When the sender is unique, energy consumption of the overhearers can be assumed quite the same as the one in case of a global buffer with one packet to send (A=1).
\begin{equation}
E_{Case_8,o}^X(2) = E_o^X(1)
\end{equation}
\end{itemize}

The overall energy cost is the sum of the costs of each scenario, weighted by the probability of the scenario to happen (as showed in the figure~\ref{fig:probabilitiesXmac}):
\begin{equation}
E^{X}(2) = \sum\limits_{i=1}^{8} p_{Case_i}\cdot E_{Case_i}^X(2)
\end{equation}


\textit{ }

\noindent\textbf{LA-MAC ($B=2$)}

\begin{figure}[h!]
\begin{center}
\includegraphics[scale=.5]{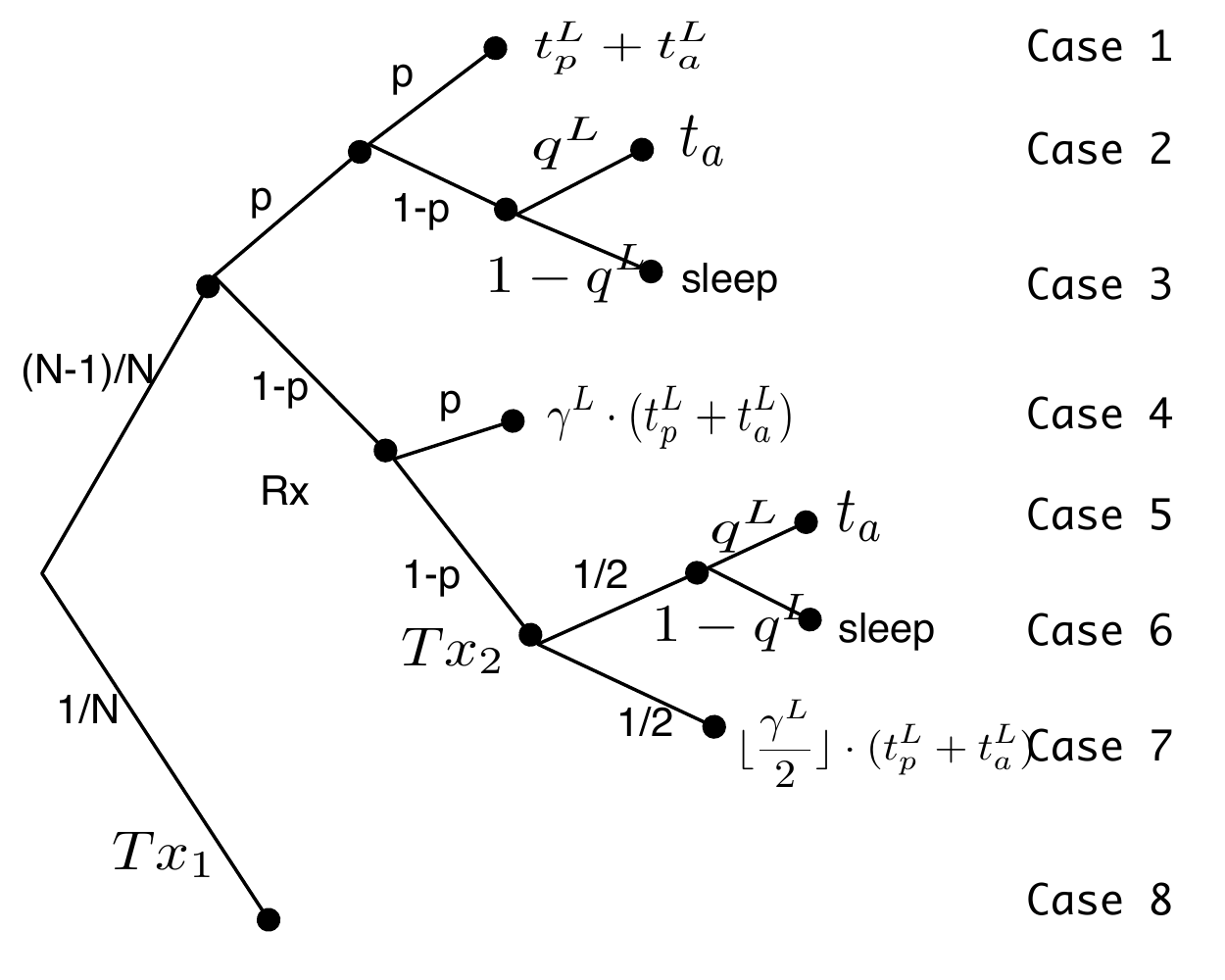}
\end{center}
\caption{LA-MAC protocol, global buffer size A=2: Probability tree of wakeup combinations.} 
\label{fig:probabilitiesLamac}
\end{figure}

When global buffer contains more than one message, there can be one or several senders. 
In this section we deal with the case $A=2$.
Energy consumption $E^L(2)$ depends on the number of senders as well as on how wake-up are scheduled.
All different combinations of wake-up instants with their probabilities are given on figure~\ref{fig:probabilitiesLamac}.
With probability $(N-1)/N$ there are two senders, otherwise there is a single sender.
Cases 1-7 refer to situations in which two senders are involved, whereas case 8 refers to a scenario with one sender.
\vspace{1cm} 

\vspace{1cm} 

\begin{itemize}
\item Case 1: The three agents are quasi-synchronized.
The very first preamble is instantly cleared by the receiver; the second transmitter hears this preamble and the ACK.
This scenario happens with a probability equal to $p_{Case_1}=(N-1)/N\cdot p\cdot p$. 
\vspace{0.5cm}

\begin{figure}[h!]
\begin{center}
\includegraphics[scale=.5]{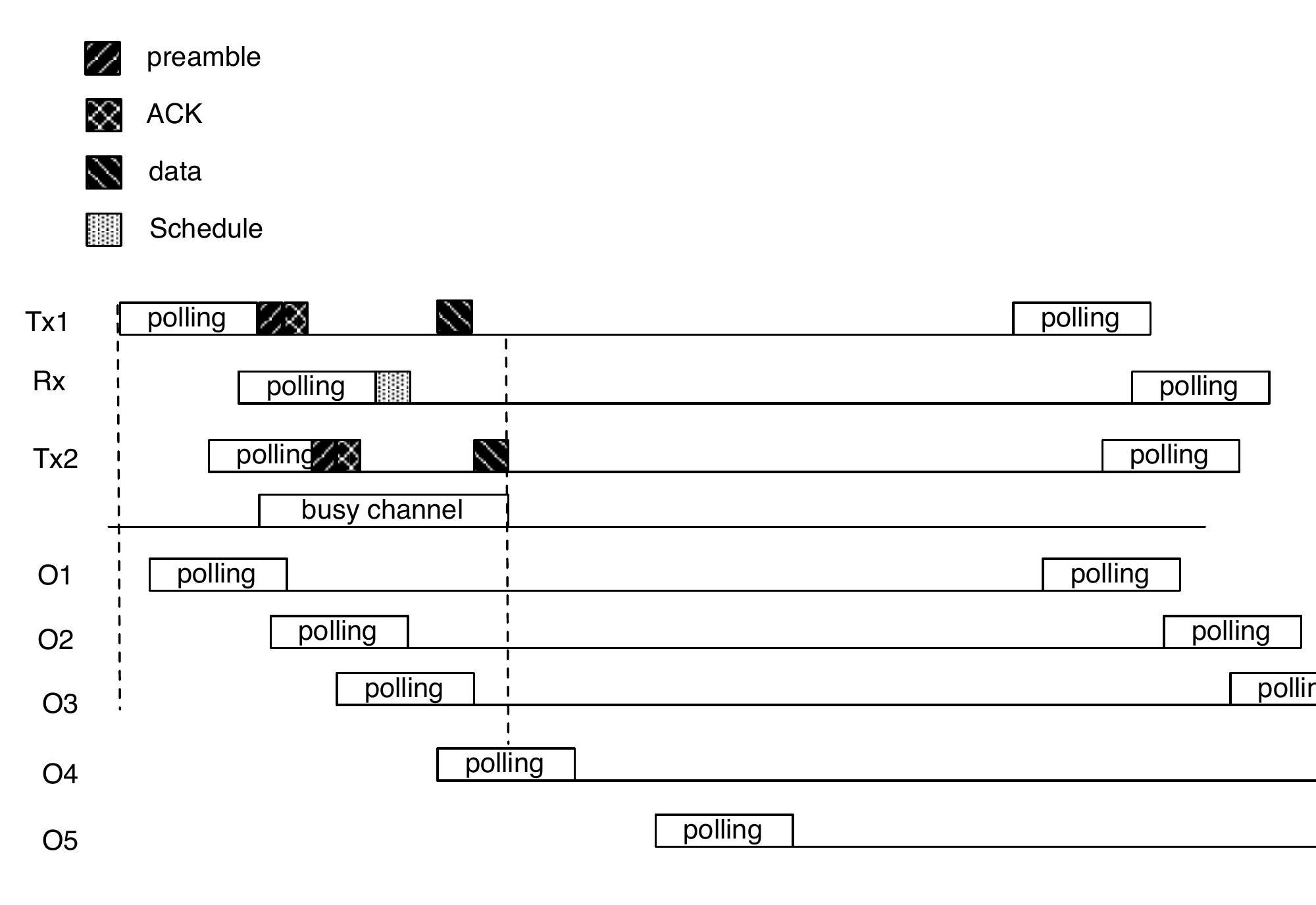}
\end{center}
\caption{LA-MAC protocol, global buffer size A=2: Overhearing situations for Case 1.}
\label{fig:lamacA2case1}
\end{figure}
\vspace{1cm}

\begin{equation}
E_{Case_1,t}^L (2)= t_p^L\cdot P_t + t_a^L\cdot P_r  + ( t_p^L  + t_a^L )\cdot  (P_r + P_t) +  t_g \cdot P_t  + t_d \cdot P_t  
\end{equation}
\begin{equation}
E_{Case_1,r}^L(2) = E_r^{L}(1) + (t_p^L + t_d) \cdot P_r   + t_a^L \cdot P_t
\end{equation}
\begin{equation}
E_{Case_1,l}^L(2) =  (E_l^{L}(1) -  (t_p^L  + t_a^L )\cdot P_l) + \frac{t_l}{2} \cdot P_l  
\end{equation}
\begin{equation}
E_{Case_1,s}^L(2) = (E_s^{L}(1) - t_d) - (t_f - \frac{t_l}{2} - t_p^L - t_a^L - t_g - t_d)\cdot P_s 
\end{equation}
As far as overhearers are concerned, several situations may happen depending on the instants of wakeup of the overhearer 
For simplicity we assume that if the overhearer is quasi-synchronized with one of the three active agents (sender one, two or receiver), it will sense a busy channel (cf. figure~\ref{fig:lamacA2case1}).
We assume that the overhearer will poll the channel for some time and then ovehear a message (that can be a preamble, an ACK, a Schedule or a data), for simplicity we assume that the ovehrearer polls the channel for an average time equal half the duration $(t_l )$ and then it will overhear a data (the largest message that can be sent). 
Probability to wakeup during a busy period is $ p_{case1,A=2}^L = (2\cdot (t_p^L +  t_a^L + t_d) + t_g ) /t_f$.
Otherwise, if the overhearer wakes up while channel is free, it will poll the channel and then go to sleep. 
\begin{equation}
E_{Case_1,o}^L(2) =  p_{case1,A=2}^L \cdot (\frac{t_l}{2} \cdot P_l +t_d \cdot P_r  + (t_f-\frac{t_l}{2} - t_d ) \cdot P_s) + (1-p_{case1,A=2}^L ) /t_f) \cdot (t_l \cdot P_l  + (t_f - t_l) \cdot P_s)
\end{equation}

\item Case 2: The first transmitter and receiver are quasi-synchronized. However the second sender, is not. The only possibility for it to send data in the current frame is to send a preamble during the polling of the receiver plus the probability to receive an ACK . This event happens with probability $q^L=1/\gamma^L + (t_l - t_a)/ t_f$.
\begin{equation}
E_{Case_2,t}^L(2) = E_t^{L}(1) +(t_g + 2 \cdot t_a^L ) \cdot P_r + ( t_p^L + t_d )\cdot P_t  
\end{equation}
\begin{equation}
E_{Case_2,r}^L(2) =   E_r^{L}(1) + (t_p^L + t_d) \cdot P_r   + t_a^L \cdot P_t
\end{equation}
\begin{equation}
E_{Case_2,l}^L(2) = (E_l^{L}(1) -  (t_p^L  + t_a^L )\cdot P_l) + \frac{t_p}{2} \cdot P_l  
\end{equation}
\begin{equation}
E_{Case_2,s}^L(2) = E_s^{L}(1) - (t_f- \frac{t_p}{2} -  t_p^L - t_a^L - t_g -  t_d)\cdot P_s 
\end{equation}
We assume that the probability of busy channel is the same as the previous case. 
So consumption is assumed to be the same as the previous case.
\begin{equation}
E_{Case_2,o}^L(2) = E_{Case_1,o}^L(2)
\end{equation}
\item Case 3: with probability $1-q^L$, the second sender wakes up too late and can not catch the acknowledge.
In this case it will go back to sleep and it will transmit its data during the next frame.
ADD FIGURE FOR THIS CASE
\begin{equation}
E_{Case_3}^L(2) =2\cdot E^{L}(1) 
\end{equation}

\item Case 4: First and second senders are quasi-synchronized but the receiver wakes up later. In this case, the first sender will send a strobed preamble and the second will hear all the preambles until the receiver wakes up and sends the ACK.
\vspace{0.5cm} 
\begin{figure}[h!]
\begin{center}
\includegraphics[scale=.5]{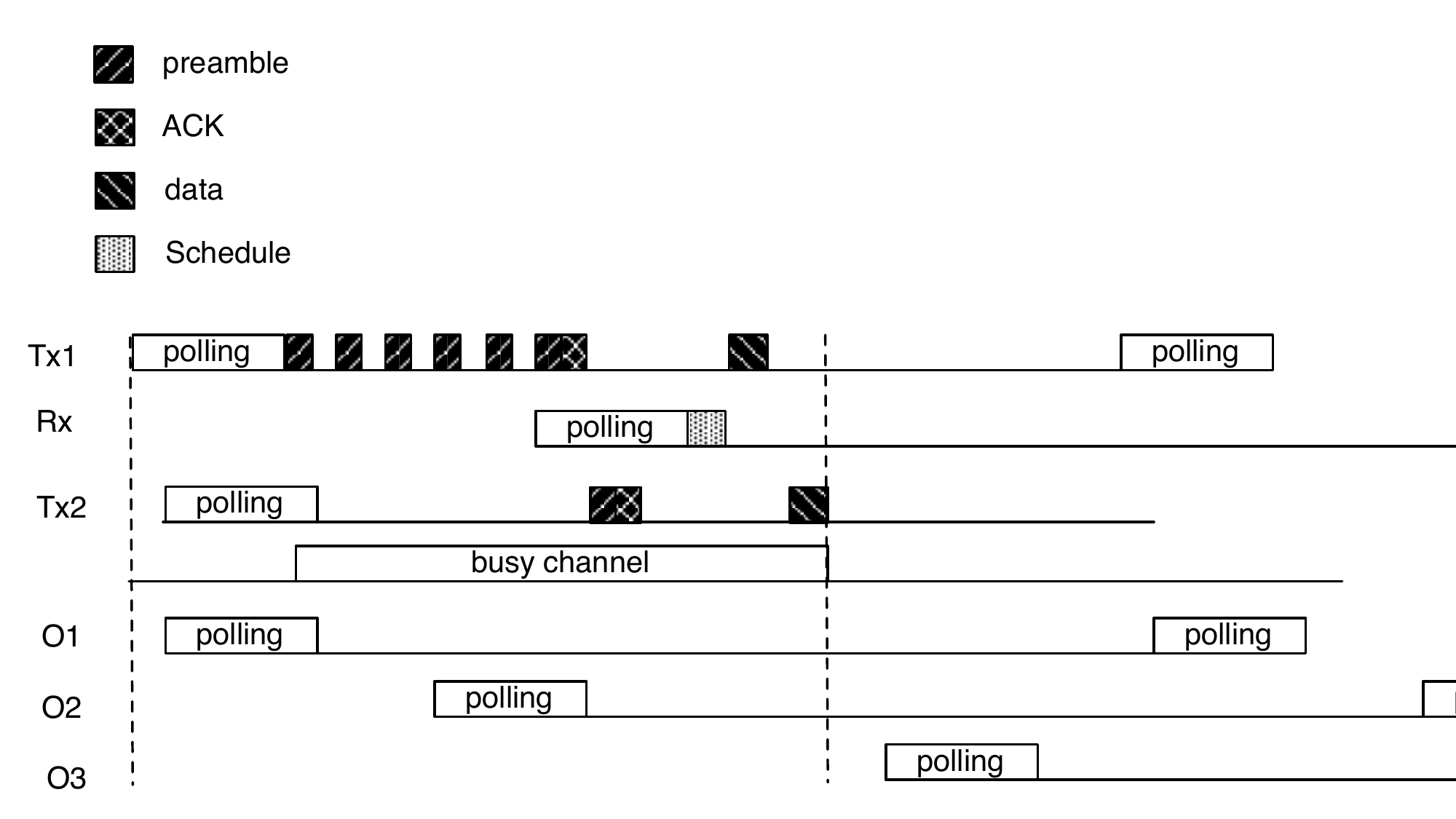}
\end{center}
\caption{Global buffer size A=2. Overhearing situations for Case 4.  LA-MAC protocol} 
\label{fig:lamacA2case4}
\end{figure}
\vspace{1cm} 
\begin{equation}
E_{Case_4,t}^L(2) = E_t^{L}(1) + \gamma^L \cdot  t_p^L   \cdot P_r + 2\cdot t_a^L\cdot P_r + t_g\cdot P_r+  (t_p^L +t_d) \cdot P_t  
\end{equation}
\begin{equation}
E_{Case_4,r}^L(2) =     E_r^{L}(1) + (t_p^L + t_d) \cdot P_r   + t_a^L \cdot P_t
\end{equation}
\begin{equation}
E_{Case_4,l}^L(2) = (E_l^{L}(1) -  (t_p^L  + t_a^L )\cdot P_l) + ((\gamma^L  -1 )\cdot   t_a^L  + \frac{ t_l}{2})\cdot P_l 
\end{equation}
\begin{equation}
E_{Case_4,s}^L(2) = E_s^{L}(1) - (t_f - (\gamma^L +1) \cdot  (  t_p^L +t_a^L  ) - t_g - t_d - \frac{ t_l}{2})\cdot P_s 
\end{equation}
If the receiver wakes up later than the couple of senders, the probability that an overhears wakes up during a transmission of a preamble is high. 
If this happens, the overhearer performs a very short polling, overhears a message (most probably a preamble) and then it goes back to sleep.
For simplicity we assume that the overhearer will perform half of $(t_p^L + t_a^L)$ of polling and than overhears an entire preamble.
Probability of busy channel is $ p_{case4}^L =\gamma^L \cdot  (  t_p^L +t_a^L  ) + (  t_p^L +t_a^L  ) + t_g + 2 \cdot t_d $

\begin{equation}
E_{Case_4,o}^L(2) = p_{case4}^L \cdot (\frac{t_p^L + t_a^L}{2} \cdot P_l + t_p^L \cdot P_r + (t_f-\frac{t_p^L + t_a^L}{2} -t_p^L )\cdot P_s ) + (1-p_{case4}^L ) \cdot (t_l \cdot P_l  + (t_f - t_l) \cdot P_s)
\end{equation}
\item Cases 5, 6, 7: The second transmitter and the receiver are not synchronized with the first transmitter, the behaviour of the protocol depends on which agent will wakes up as first among the second transmitter and the receiver.   
\begin{itemize}
\item Case 5: the receiver wakes up as first, similarly to Case 2, the only possibility for it to send data in the current frame is to listen to the ACK of the receiver, during its polling. This event happens with probability $q^L=1/ \gamma^L$.
\vspace{0.5cm} 
\begin{figure}[h!]
\begin{center}
\includegraphics[scale=.5]{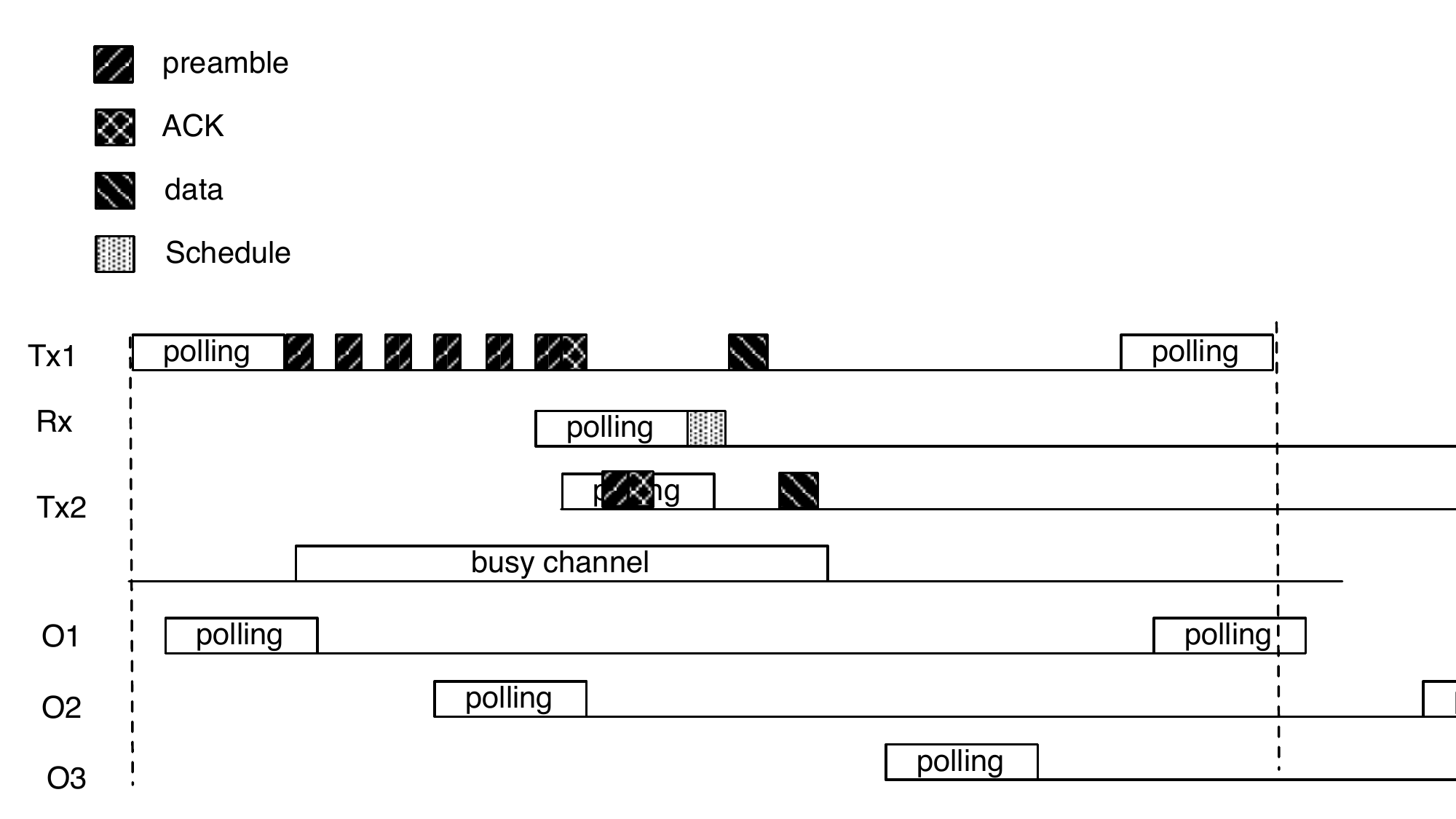}
\end{center}
\caption{Global buffer size A=2. Overhearing situations for Case 5.  LA-MAC protocol} 
\label{fig:lamacA2case5}
\end{figure}
\vspace{1cm} 
\begin{equation}
E_{Case_5,t}^L(2) = E_{Case_2,t}^L
\end{equation}
\begin{equation}
E_{Case_5,r}^L(2)  = E_{Case_2,r}^L 
\end{equation}
\begin{equation}
E_{Case_5,l}^L(2) = E_{Case_2,l}^L 
\end{equation}
\begin{equation}
E_{Case_5,s}^L(2) = E_{Case_2,s}^L 
\end{equation}
As in the previous case, the overhearer will perceive a very busy channel because of the transmission of preambles so when it will wake up, it  will perform half of $(t_p^L + t_a^L)$ in polling state  and than it will overhear an entire preamble.
Probability of busy channel is $ p_{case5}^L =p_{case4}^L =\gamma^L \cdot  (  t_p^L +t_a^L  ) + (  t_p^L +t_a^L  ) + t_g + 2 \cdot t_d $

\begin{equation}
E_{Case_5,o}^L(2) =E_{Case_4,o}^L(2) 
\end{equation}

\item Case 6: the receiver wakes up as first, similarly to Case 3, with probability $1-q^L$, the second sender wakes up too late and can not catch the acknowledge. In this case it will go back to sleep and it will transmit its data during the next frame.
\begin{equation}
E_{Case_6}^L(2) = E_{Case_3}^L =2\cdot E^{L}(1) 
\end{equation}

\item Case 7: the second transmitter wakes up as first, will hear a part of the strobed preamble until the receiver wakes up and sends the ACK. 
In average, the second transmitter will hear $\lfloor \dfrac{\gamma^L}{2}\rfloor$ preambles.
\begin{equation}
E_{Case_7,t}^L(2) = E_t^{L}(1) + \lfloor\frac{\gamma^L}{2}\rfloor \cdot  t_p^L   \cdot P_r +2 \cdot t_a^L \cdot P_r + (t_p^L+  t_d) \cdot P_t  + t_g \cdot P_r
\end{equation}
\begin{equation}
E_{Case_7,r}^L(2) =     E_r^{L}(1) + (t_p^L + t_d) \cdot P_r   + t_a^L \cdot P_t
\end{equation}
\begin{equation}
E_{Case_7,l}^L(2) =  (E_l^{L}(1) -  (t_p^L  + t_a^L )\cdot P_l) +((\lfloor\frac{\gamma^L}{2}\rfloor  -1 )\cdot   t_a^L  +  \frac{t_p^L + t_a^L}{2})\cdot P_l 
\end{equation}
\begin{equation}
E_{Case_7,s}^L(2) = E_s^{L}(1) - (t_f - ((\lfloor\frac{\gamma^L}{2}\rfloor +1) \cdot  (  t_p^L +t_a^L  )- \frac{t_p^L + t_a^L}{2}-t_g - t_d)\cdot P_s 
\end{equation}
From the overhearers point of view, this case is equivalent to Cases 4 and 5.
\begin{equation}
E_{Case_7,o}^L(2) = E_{Case_4,o}^L(2)
\end{equation}
\end{itemize}

\item Case 8: there is only one sender that will send two messages in a row.

\begin{equation}
E_{Case_8,t}^L(2) = E_t^{L}(1) +  t_d \cdot P_t  
\end{equation}
\begin{equation}
E_{Case_8,r}^L(2) =    E_r^{L}(1) + t_d \cdot P_r  
\end{equation}
\begin{equation}
E_{Case_8,l}^L(2) =  (E_l^{L}(1) - t_d \cdot P_l) 
\end{equation}
\begin{equation}
E_{Case_8,s}^L(2) = E_s^{L}(1) - t_d \cdot P_s 
\end{equation}
When the sender is unique, overhearer consumption can be assumed the same as the case of A=1.
\begin{equation}
E_{Case_8,o}^L (2)= E_o^L(1)
\end{equation}
\end{itemize}

The overall energy cost is the sum of the costs of each case weighted by the probability of the case to happen (as showed in the figure~\ref{fig:probabilitiesXmac}):
\begin{equation}
E^{L}(2) = \sum\limits_{i=1}^{8} p_{Case_i}\cdot E_{Case_i}^L
\end{equation}

\subsection{Global buffer contains more than two messages ($B>2$)}


\textit{ }

\noindent\textbf{B-MAC ($B>2$)}


\textit{ }

\noindent\textbf{X-MAC ($B>2$)}


\textit{ }

\noindent\textbf{LA-MAC ($B>2$)}

\begin{figure}[ht!] 	
	\begin{center} 	
	\includegraphics[scale=.7]{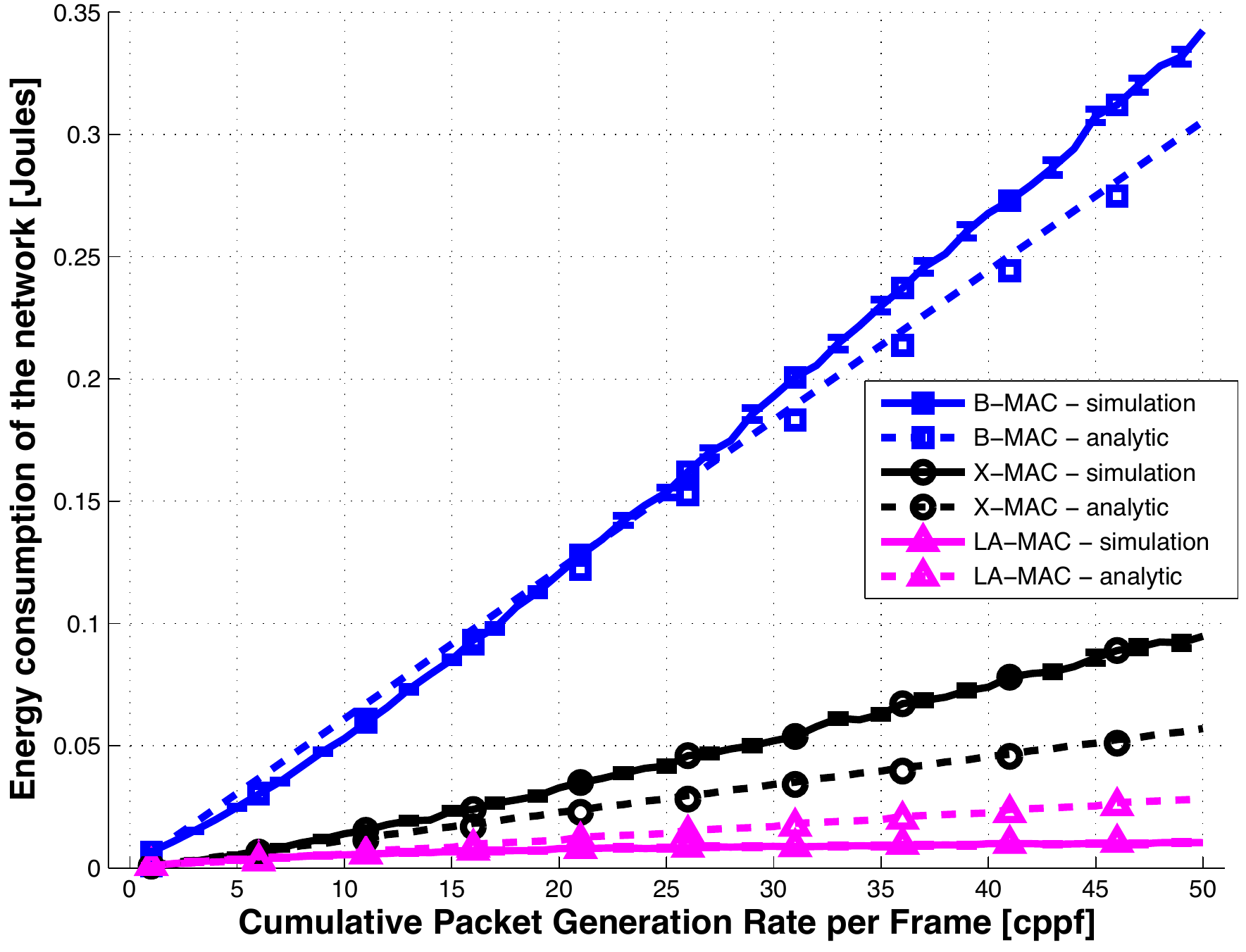} 	 		
	\caption{Energy analysis and OMNeT++ simulations versus the global buffer size.} 	
	\label{fig:expected_energy_consumption} 	
	\end{center} 	
\end{figure}

\section{Numerical Validation}
\label{sec_simulation}
We have implemented the analyzed MAC protocols in the  OMNeT++ simulator
\cite{omnetpp.simulator} for numerical evaluation.  
Each numerical value is the average of 100 runs and we show the
corresponding confidence intervals at 95$\%$ confidence level. 
We assume that devices use   the
CC1100~\cite{chipcon.cc1100} radio stack with bitrate of 20Kbps. 
The values of power consumption for different radio states are specific to the
CC1100
transceiver considering a 3V battery.  
In the following, we assume $N=9$ senders.
The periodical wakeup period is the same for all protocols: $t_f = t_l + t_s =
250~ms$. 
Also the polling duration is the same for all protocols: $t_l=25~ms$, thus the
duty cycle with no messages to send is $10\%$.
We provide numerical and analytical results for buffer size $B \in[1,50]$. 
We compare the protocol performance with respect to several criteria:

\begin{itemize}
\item\textit{Latency [s]:} the delay between the beginning of the simulation 
and the instant of packet reception at the sink (we present the latency averaged
over all nodes).

\item\textit{Energy Consumption [Joules]:} the averaged energy consumed by all nods due to
radio activity.

\item\textit{Delivery Ratio:} the ratio of the number of received packet by
the sink to the total number of packets sent.
\end{itemize}

\begin{figure}[ht!] 	
    	\begin{center} 	
   	\includegraphics[scale=.7]{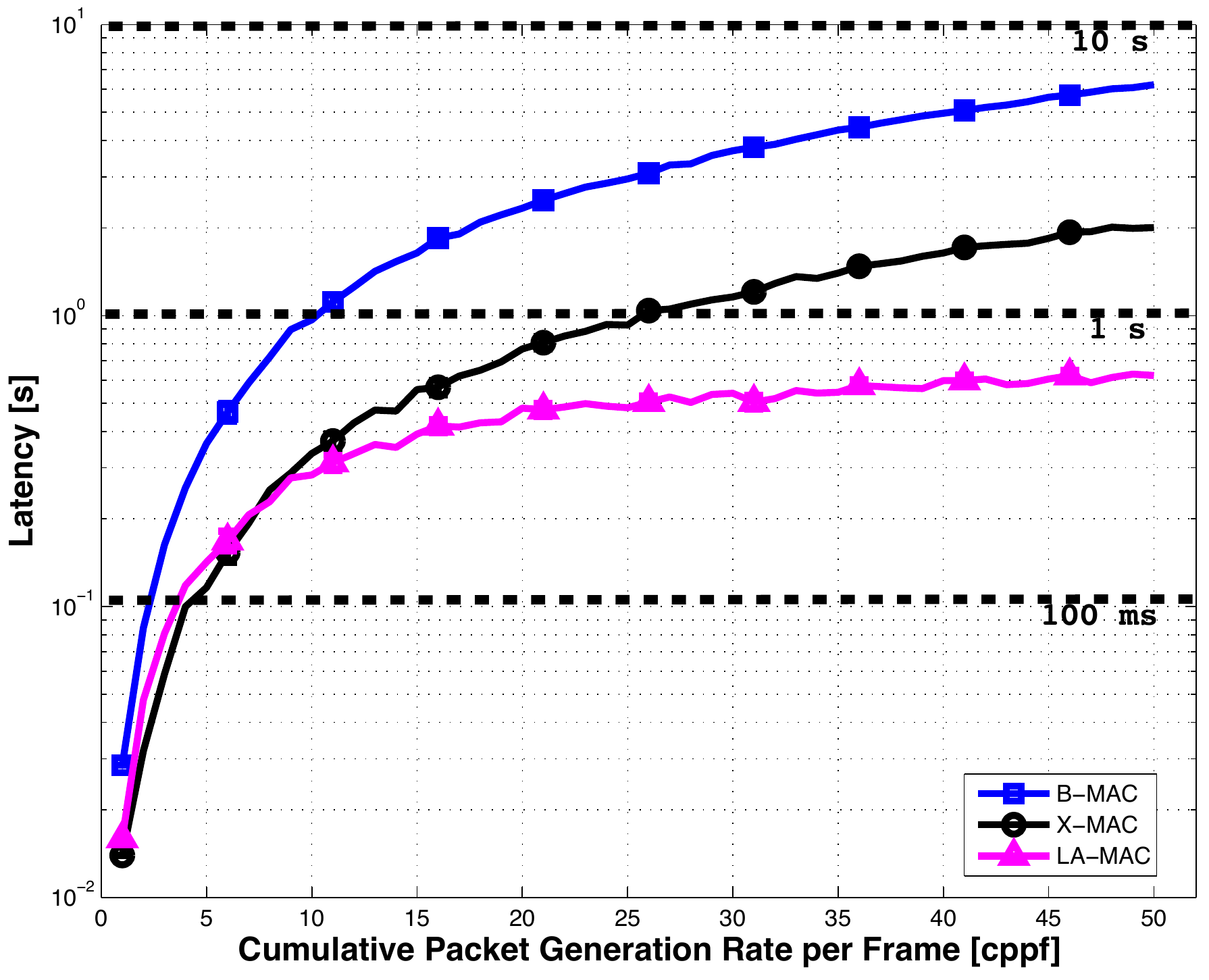} 		
   	\caption{Average latency versus the global buffer size.} 		
   	\label{fig:latency}
   	\end{center} 	
\end{figure}

\begin{figure}[ht!] 	
      	\begin{center} 	
     	\includegraphics[scale=.7]{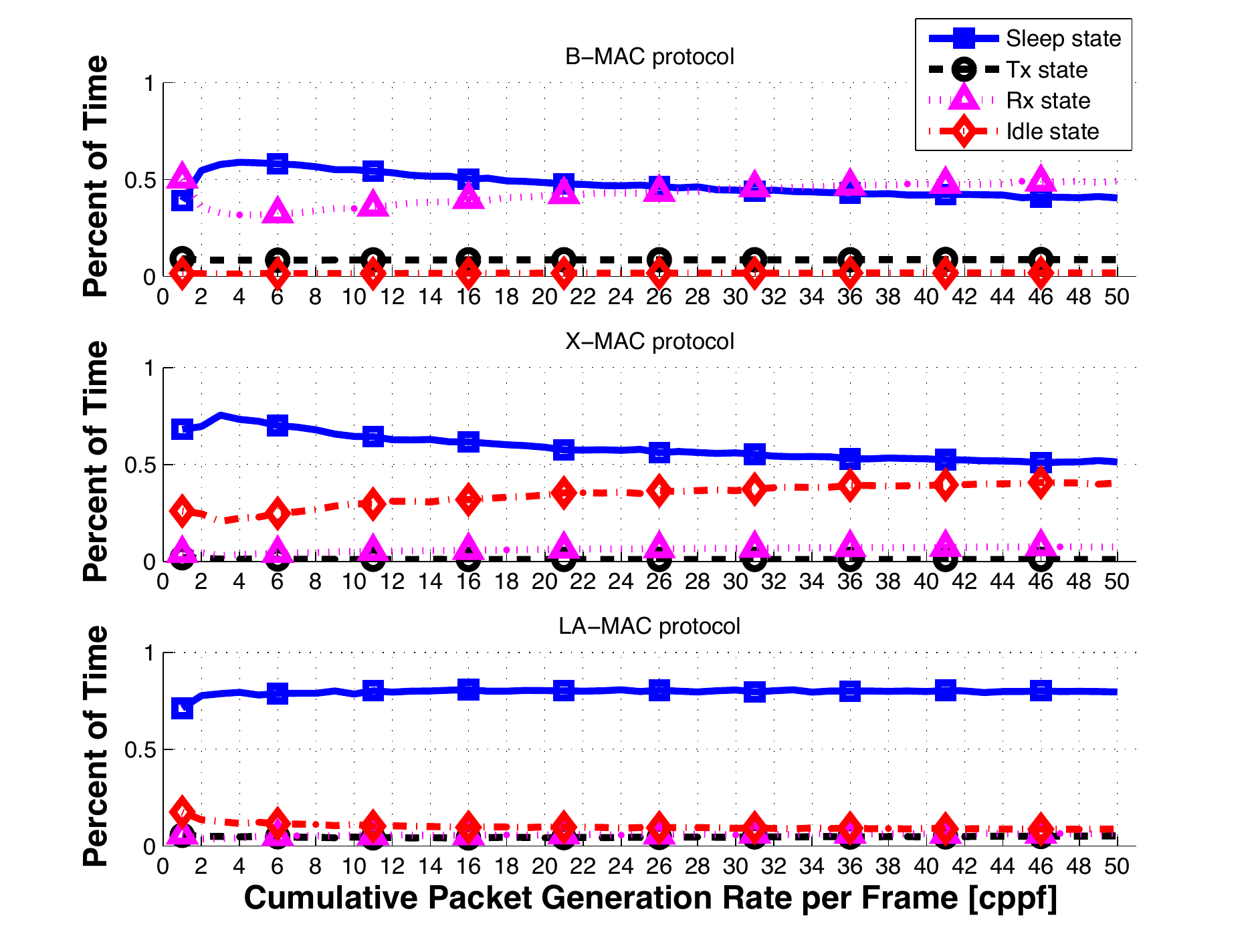} 		
     	\caption{Percentage of the time spent in each radio state versus the
          global buffer size.}
     	\label{fig:radio_states}
     	\end{center} 	
\end{figure}

In Figure~\ref{fig:expected_energy_consumption}, we show the comparison between
the proposed energy consumption analysis and numerical simulations for different
values of the global buffer size. 
We assume that at the beginning of  each simulation all messages to send are
already buffered.
Each simulation stops when the last message in the buffer is received by the
sink.
Figure~\ref{fig:expected_energy_consumption} highlights the validity of  the
analytical expressions for energy consumption: all curves match very well.
As expected, B-MAC is the most energy consuming protocol: as the buffer size
increases, the transmission of a long preamble locally saturates the network
resulting in high energy consumption and latency (cf. Figure~\ref{fig:latency}).
\begin{figure}[ht!] 
	\begin{center} 		
	\includegraphics[scale=.7]{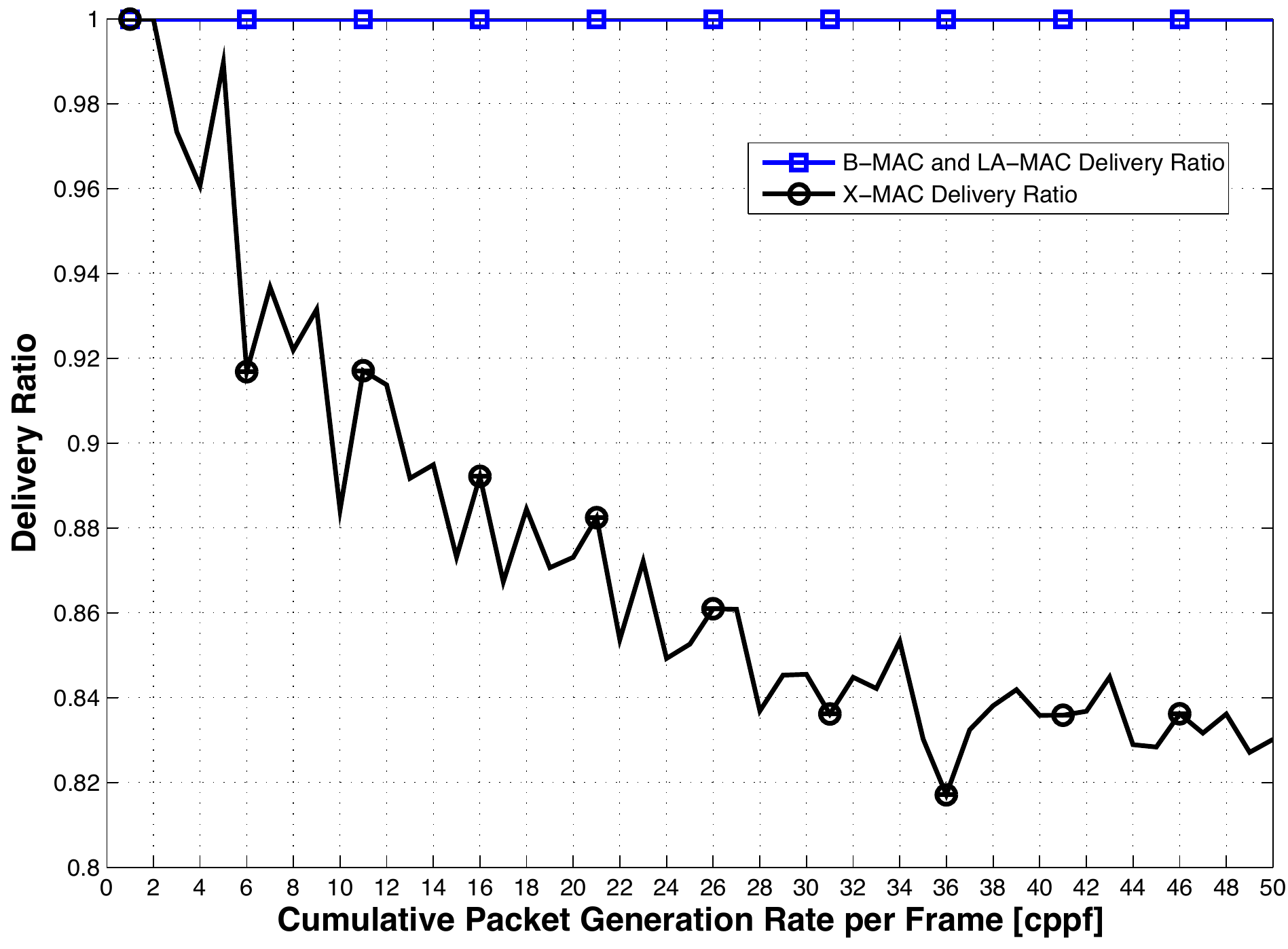} 		
	\caption{Delivery ratio versus the global buffer size. In X-MAC, most collisions
happen when messages
are sent after the back-off time. } 		
	\label{fig:deliv} 	
	\end{center} 
\end{figure}  
In X-MAC, short preambles mitigate the effect of the increasing local traffic
load, 
thus both latency and energy consumption are reduced with respect to B-MAC.
Even if X-MAC is more energy efficient than B-MAC, Figure~\ref{fig:deliv} shows
that even for small buffer sizes, the delivery ratio for this protocol is lower
than 100 $\%$ most likely because packets that are sent after the back-off
collide at the receiver.
LA-MAC is the most energy saving protocol and it also outperforms other
protocols in terms of latency and  the delivery ratio.
We observe that when the instantaneous buffer size is lower than 8 messages, the
cost of the SCHEDULE message is paid in terms of a higher latency with respect to
X-MAC (cf. Figure~\ref{fig:latency}); however, for larger buffer sizes the cost
of the SCHEDULE transmission is compensated by a high number of delivered
messages.
In Figure~\ref{fig:radio_states}, we show the percentage of the time during
which devices spend in each radio state versus the global buffer size. 
Thanks to efficient message scheduling of LA-MAC, devices sleep most of the time
independently of the buffer size and all messages are
delivered.

\section{Conclusions} 
\label{sec:conclusion}
In  the present paper, we have analyzed the energy
consumption of preamble sampling MAC protocols by means of a simple
probabilistic modeling. The analytical results are then 
validated by simulations.
We compare the classical MAC protocols (B-MAC and X-MAC) with LA-MAC, a method
proposed in a companion paper. 
Our analysis highlights the energy savings achievable with LA-MAC with respect
to B-MAC and X-MAC. It also shows that
LA-MAC provides the best performance in the considered case of high density
networks under traffic congestion.



\bibliographystyle{IEEEtran}
\bibliography{bibfiles/macbib,bibfiles/WSNapplications,bibfiles/publiabbrev,bibfiles/BIBinfocom,bibfiles/biblio-wcds}

\end{document}